\documentclass[12pt]{article}

\usepackage[normalem]{ulem}

\usepackage{xcolor}
\usepackage[english]{babel}
\usepackage[T1]{fontenc}
\usepackage[utf8]{inputenc}
\usepackage{authblk}
\usepackage{mathtools}
\usepackage{slashed}
\usepackage{amsmath,amssymb}
\usepackage{amsfonts}
\usepackage{graphicx,color}
\usepackage{cite}
\usepackage{hyperref}
\hypersetup{
    colorlinks=true,
    linkcolor=blue,
    filecolor=magenta,      
    urlcolor=cyan,
}
\usepackage{cleveref}

\numberwithin{equation}{section} 

\definecolor{refcol}{rgb}{0.9,0.1,0.1}
\hypersetup{colorlinks=true,linkcolor=blue,citecolor=refcol,urlcolor=cyan,linktocpage}

\newcommand{\ben}{\begin{eqnarray}\displaystyle}
\newcommand{\een}{\end{eqnarray}}

\newcommand{\be}{\begin{equation}}
\newcommand{\ee}{\end{equation}}

\newcommand{\bc}{\begin{center}}
\newcommand{\ec}{\end{center}}

\newcommand{\eesp}{\end{split}}
\newcommand{\bsp}{\begin{split}}

\newcommand{\Rmnum}[1]{\expandafter\@slowromancap\romannumeral #1@}

\newcommand{\cH}{\mathcal{H}}

\newcommand{\cM}{\mathcal{M}}

\newcommand{\cS}{\mathcal{S}}

\newcommand{\gbar}{\bar{g}}
\newcommand{\Abar}{\bar{A}}
\newcommand{\Fbar}{\bar{F}}

\newcommand{\bensp}{\begin{eqnarray}\begin{split}}
\newcommand{\eensp}{\end{eqnarray}\end{split}}

\newcommand{\bnm}{\begin{matrix}}
\newcommand{\enm}{\end{matrix}}

\def\XXint#1#2#3{{\setbox0=\hbox{$#1{#2#3}{\int}$ }
\vcenter{\hbox{$#2#3$ }}\kern-.6\wd0}}

\newcommand{\nhzeta}{\zeta_{\text{nh}}}
\newcommand{\bzeta}{\zeta_{\text{bdy}}}
\newcommand{\mO}{\mathcal{O}}

\newcommand{\cb}{\mathfrak{b}}
\newcommand{\cg}{\mathfrak{g}}
\newcommand{\ch}{\mathfrak{h}}

\renewenvironment{quote}{%
  \list{}{%
    \leftmargin0.3cm   
    \rightmargin\leftmargin
  }
  \item\relax
}
{\endlist}

\begin{document}

\begin{titlepage}
\thispagestyle{empty}

\title{
{\Huge\bf Flow of shear response functions in hyperscaling violating Lifshitz theories}
}

\vfill

\author{
    {\bf Arghya Chattopadhyay}${}^{1}$\thanks{{\tt arghya.chattopadhyay@wits.ac.za}},\hspace{8pt}
    {\bf Nihal M}\thanks{{ \tt Majority of the work was done while the author was affiliated to:}\\ {\it  School of Physics, IISER Thiruvananthapuram, Thiruvananthapuram - 695551, India.}}{\hspace{0.2cm}}\thanks{{\tt nihalmh9816@alumni.iisertvm.ac.in}}
    ,\hspace{8pt}
    {\bf Debangshu Mukherjee}${}^2$\thanks{{\tt debangshu@iitk.ac.in}}
\hfill\\
{\small
${}^1${\it National Institute of Theoretical and Computational Sciences,}\\
{\it School of Physics and Mandelstam Institute for Theoretical Physics}\\
{\it University of the Witwatersrand, Wits, 2050, South Africa.}\\

\smallskip\hfill\\
${}^2${\it Department of Physics, Indian Institute of Technology Kanpur, }\\
{\it  Kanpur 208016, India}
 }
     }
\vfill

\date{
\begin{quote}
\centerline{{\bf Abstract}}
\smallskip\hfill\\
{\small We study the flow equations of the shear response functions for hyperscaling
violating Lifshitz (hvLif) theories, with Lifshitz and hyperscaling violating
exponents $z$ and $\theta$. Adapting the membrane paradigm approach of
analysing response functions as developed by Iqbal and Liu, we focus
specifically on the shear gravitational modes which now are coupled to the
perturbations of the background gauge field. Restricting to the zero momenta
sector, we make further simplistic assumptions regarding the hydrodynamic
expansion of the perturbations. Analysing the flow equations shows that the shear viscosity at leading order saturates the Kovtun-Son-Starinets (KSS) bound of $\frac{1}{4\pi}$. When $z=d_i-\theta$, ($d_i$ being the number of spatial dimension in the dual field theory) the first-order correction to shear viscosity exhibits logarithmic scaling, signalling the emergence of a scale in the UV regime for this class of hvLif theories. We further show that the response function associated to the gauge field perturbations diverge near the boundary when $z>d_i+2-\theta$. This provides a holographic understanding of the origin of such a constraint and further vindicates results obtained in previous works that were obtained through near horizon and quasinormal mode analysis.}
\end{quote}
}

\vfill

\end{titlepage}
\thispagestyle{empty}\maketitle\vfill \eject

\tableofcontents

\section{Introduction}
\label{sec:intro}

The framework of gauge/gravity duality \cite{Maldacena:1997re,Gubser:1998bc,Witten:1998qj,Aharony:1999ti} has been generalized and applied to understand strongly coupled non-relativistic field theories. In particular, a certain class of non-relativistic field theories, dubbed as hyperscaling violating Lifshitz (hvLif) theories (which are conformal to Lifshitz theories) has been extensively explored in previous works \cite{Gubser:2009qt,Cadoni:2009xm,Goldstein:2009cv,Charmousis:2010zz,Perlmutter:2010qu,Bertoldi:2010ca,Kim:2010zq,Iizuka:2011hg,Ogawa:2011bz,Cremonini:2011ej,Huijse:2011ef,Dong:2012se,Kiritsis:2012ta,Bhattacharya:2012zu,Alishahiha:2012cm,Hoyos:2013qna,Sadeghi:2014zia,Ghodrati:2014spa,Kiritsis:2015doa,Kuang:2015mlf,Taylor:2015glc,Kolekar:2016pnr,Kiritsis:2016rcb,Kolekar:2016yzg,Ling:2016ien,Ling:2016yxy,Davison:2016auk,Hartong:2016nyx,Eberlein:2016luy,Mukherjee:2017ynv,Hartnoll:2018xxg,Herrera-Aguilar:2021top,Yuan:2020fvv}. In fact, there are concrete examples of realizable condensed matter systems where certain correlators exhibit similar scaling behaviour as that of hvLif theories\cite{Hartnoll:2018xxg}. Interested readers can see \cite{Taylor:2015glc,Hartnoll:2018xxg} for a comprehensive review of these class of non-relativistic field theories. 

The gravity dual of hvLif theories can be realized as solutions to effective Einstein-Maxwell-dilaton theories\cite{Gubser:2009qt,Cadoni:2009xm,Goldstein:2009cv,Charmousis:2010zz,Perlmutter:2010qu,Bertoldi:2010ca,Kim:2010zq,Iizuka:2011hg,Ogawa:2011bz,Cremonini:2011ej,Huijse:2011ef,Dong:2012se,Kiritsis:2012ta,Bhattacharya:2012zu,Alishahiha:2012cm}. hvLif solutions may be embedded in string theory as null reductions of boosted black branes \cite{Narayan:2012hk,Singh:2012un} (Lifshitz spacetimes which are conformal to hvLif spacetimes also admit gauge/string realizationss\cite{Balasubramanian:2010uk,Donos:2010tu,Ross:2011gu,Christensen:2013rfa,Chemissany:2014xsa,Hartong:2015wxa}). For a better understanding of this class of non-relativistic field theories, it is crucial to understand their infrared (IR) behaviour, in particular, hydrodynamics and various response functions that emerges in the low-energy limit. In previous works, the shear diffusion constant and the shear viscosity bound for hvLif theories were analysed using the membrane paradigm approach\cite{Kolekar:2016pnr} as well as quasi-normal modes of the dual gravity theory \cite{Mukherjee:2017ynv}. It was found that for a $d_i+1$-dimensional hvLif theory with Lifshitz exponent $z$ and hyperscaling violating exponent $\theta$, one must have $z \leq d_i+2-\theta$ for a consistent hydrodynamic expansion. When $z=d_i+2-\theta$, the shear diffusion constant exhibits a novel logarithmic scaling while the Kovtun-Starinets-Son (KSS) shear viscosity bound is saturated\cite{Kovtun:2004de}. For $z >d_i+2-\theta$, the first order solution diverges at the boundary presumably hinting towards a breakdown of the hydrodynamic expansion for this parameter regime.

In this paper, we take the approach as pioneered by Iqbal and Liu \cite{Iqbal:2008by}. The gauge/gravity duality maps the strongly coupled field theory on the boundary to the weakly coupled black hole spacetime in the bulk. However, the membrane paradigm approach to black holes endows hydrodynamic properties such as viscosity, entropy, conductivity etc. to a fictitious \emph{stretched horizon} which is hovering very close to the real event horizon. Using the UV/IR point of view, Iqbal and Liu essentially attempted to relate this horizon fluid to the hydrodynamic regime of the strongly coupled field theory living on the boundary in the context of $AdS$ gravity. It turned out that in the low-frequency, long wavelength limit (i.e. hydrodynamic limit) the evolution of retarded Green's function of the boundary with respect to energy scale is trivial. To be more precise, one can think of the radial direction of the bulk gravity theory as the energy scale of the boundary theory. Thus, the perturbed bulk Einstein's equations at linearized order can be thought of as a RG flow equation for a certain generalized response function which turns out to be independent of the radial direction at leading order. The \emph{triviality} of flow of the response function implies that the corresponding transport coefficient can be expressed in terms of geometric quantities over any constant $r$ hypersurface of the bulk theory and hence can be shown to be \emph{universal}.

The aim of this work is to adapt the above approach and study the RG flow of response functions in the context of hvLif theories. The analysis is significantly more complicated due to nontrivial coupling between the shear perturbative modes with the gauge field perturbations. This is to be contrasted with previous works such as \cite{Mamo:2012sy,Ghodrati:2014spa} where such flow equations were studied in the context of anisotropic gravity duals or the background resulted from higher derivative corrected action. In such cases, the holographic duals interpolate between Lifshitz or hvLif in the deep IR while it asymptotes to pure $AdS$ near the boundary. 

The starting point of our analysis is a $(d+1)$-dimensional gravity dual of hvLif theory. Turning on perturbations of the form $e^{-i\omega t+iqx}h_{\mu \nu}(r)$ and $e^{-i\omega t+iqx}a_{\mu}(r)$ respectively for the metric and gauge field, we notice the shear sector modes $h_{xi}, h_{ti}$ and $a_i$ (where $i$ runs over all boundary direction except $t$ and $x$) forms a coupled set of differential equations. We associate a conjugate momenta to each of these perturbation modes and correspondingly define appropriate response functions. As one would expect, the radial flow equations for each of these response functions also follow complicated coupled non-linear differential equations. However, one must note that our principal aim is to extract the transport coefficient out of these response function in the hydrodynamic limit which one does in the language of linear response theory, adapted to the context of gauge/gravity duality. Consider a generic field theory containing an operator $\mathcal{O}$ which is coupled to a source $\varphi$. At the level of linear response they are related as
\begin{equation}
	\langle \mathcal{O}(\omega, q) \rangle= -G^{R}(\omega, q)\varphi(\omega, q)
\end{equation}
where $\omega$ and $q$ are very small frequency and momenta respectively, while $G^{R}$ denotes the retarded correlator for the operator $\mathcal{O}$. The corresponding transport coefficient is defined as
\begin{equation}
	\label{eq:chi-defn}
	\chi= \lim_{\omega \to 0} \frac{G^R(\omega, q=0)}{i\omega}\ , 
\end{equation}  
which is known as Kubo's formula. In particular, when $\mathcal{O} \equiv T^{xy}$, the corresponding transport coefficient is the shear viscosity $\eta$ while for a charge current i.e. $\mathcal{O} \equiv J^{x}$, the analogous transport coefficient is the DC conductivity. Since in the above we essentially require to find the response function at zero momenta, we focus on that regime and analyse the flow equations. Interestingly, we see that indeed for $q=0$, the flow equation for $\chi_{xi}$ i.e. the response function corresponding to $h_{xi}$ follows a Riccati equation which leads to a constant $\chi_{xi}$ at leading order for all values of $z$ and $\theta$. This behaviour is identical to that encountered in pure $AdS$ gravity\cite{Iqbal:2008by}. However, when $z=d_i-\theta$, the first order correction to the response function has a logarithmic scaling which diverges at the boundary $r \rightarrow 0$. This necessitates the introduction of a cut-off presumably signifying the UV scale beyond which the hydrodynamic expansion breaks down. 

The analysis for the response function associated with $a_i$ i.e. $\chi_{a_i}$ is more involved due to the complicated nature of the flow equation. In fact at $q=0$, it turns out the variable $\zeta_{a_i}=\omega \chi_{a_i}$ seems to admit a hydrodynamic expansion. In order to analyse the behaviour of $\zeta_{a_i}$, we focus on the near-horizon region and the near-boundary region separately which somewhat simplifies the analysis. At leading order itself, we see the solutions for $\zeta_{a_i}$ are different for the two different regimes. This is different qualitatively from the behaviour of $\chi_{xi}$ which followed a \emph{trivial flow} equation allowing one to write the response function at any point along the radial direction. Interestingly, we see close to the boundary, the leading piece of $\chi_{a_i}$ diverges when $z> d_i+2-\theta$ which is identical to results obtained in earlier works\cite{Kolekar:2016pnr,Kolekar:2016yzg,Mukherjee:2017ynv}. To further vindicate our result, one can look at the \emph{Markovianity} index of the fluctuating modes in the spirit of \cite{Ghosh:2020lel}. Interestingly, we observe that for $z\leq d_i+2-\theta$, the fluctuations starts to behave like a non-Markovian probe.

The paper is organized as follows: In \cref{sec:2}, we describe our setup and define appropriate response functions corresponding to the shear gravitational modes. The general flow equations are worked out which describes the non-perturbative evolution of response function for arbitrary frequency and momenta. \Cref{sec:3} focuses on the zero momenta sector and look at the transport coefficient associated with the modes $h_{xi}, h_{ti}$ and $a_i$. Finally, keeping some details of the calculations in three appendices, we end with a discussions of our main results with possible future directions along with a simple analysis of the \emph{Markovianity index} results in \cref{sec:4}. 
  
\section{Flow equations of response functions}\label{sec:2}

We are considering a hvLif theory living in $d=d_i+1$ spacetime dimensions with Lifshitz exponent $z$ and hyperscaling violating exponent $\theta$. This field theory has a $(d+1)$-dimensional gravity dual given by
\begin{equation}\label{hvmetric}
ds^2=r^{\frac{2\theta}{d_i}}\left(-\frac{f(r)}{r^{2z}}dt^2+\frac{dr^2}{r^2f(r)}+\sum_{i=1}^{d_i}\frac{dx_i^2}{r^2}\right)\ ; \quad d=d_i+1\ ; \quad f(r)=1-(r_0r)^{d_i+z-\theta}\ .
\end{equation}
The above metric is a solution to Einstein-Maxwell-dilaton theory (details of background solution in \cref{hvlif-review}). The temperature of the field theory dual to the hvLif theory \eqref{hvmetric} is the Hawking temperature of the black brane
\begin{equation}
\label{eq:temperature}
    T= \frac{d_i+z-\theta}{4\pi}r_0^z\ ,
\end{equation}
where the event horizon is located at $r=\frac{1}{r_0}$. 

As per the holographic dictionary, the radial coordinate $r$ can be thought of as the energy scale in the bulk theory. Our central goal in this section is to essentially set up the RG flow equations governing the response functions that we want to study. In order to obtain the RG flow equations, we turn on linearized perturbations in the bulk theory, which in general is given as,
\begin{equation}
g_{\mu \nu}=\gbar_{\mu \nu}+h_{\mu \nu}\ ; \quad A_{\mu}=\Abar_{\mu}+a_{\mu}\ ; \quad \phi = \bar{\phi} + \varphi\ ,
\end{equation}
where quantities $\bar{g}_{\mu \nu}, \bar{A}_{\mu}$ and $\bar{\phi}$ denote background fields as given in \cref{hvlif-review}. We turn on perturbations of the form $e^{-i\omega t+i qx}h_{\mu \nu}(r)$ and $e^{-i\omega t+i qx}a_{\mu}(r)$ and restrict ourselves to the radial gauge ($h_{\mu r}=a_r=0$). The shear gravitation modes $h_{xi}$ now couples to $h_{ti}$ and $a_i$ where the index $i$ runs over all boundary coordinates except $t$ and $x$. For convenience, we define the following field variables
\begin{equation}\label{h-redefn}
H_{xi}=g^{ii}h_{ix}=r^{2-\frac{2\theta}{d_i}}h_{xi}\quad ;\quad H_{ti}=g^{ii}h_{ti}=r^{2-\frac{2\theta}{d_i}}h_{ti}\ .
\end{equation}
In terms of these modes the equations of motion take the form
\begin{eqnarray}\label{nu-t-eqn}
\partial_r(r^{z+\theta-(d_i+1)}H'_{ti})-ka'_i -\frac{r^{z+\theta-(d_i+1)}}{f}\ q(\omega H_{xi}+q H_{ti})&=&0 ,\\
\label{nu-x-eqn}
\partial_r(r^{\theta-z-d_i+1}fH'_{xi})+\frac{r^{z+\theta-(d_i+1)}}{f}\ \omega (\omega H_{xi}+q H_{ti})&=&0\ ,\\
\label{nu-r-eqn}
qr^{2-2z}H'_{xi}+\frac{\omega}{f}(H'_{ti}-kr^{(d_i+1)-z-\theta}a_i)&=&0 ,\\
\label{chi-eqn}
\partial_r(r^{d_i+3-z-\theta}f a'_i)+\frac{r^{d_i+1+z-\theta}}{f}\omega^2 a_i -r^{d_i+3-z-\theta}q^2a_i-kH'_{ti}&=&0\ ,
\end{eqnarray}
where $k=(d_i+z-\theta)\alpha$. The above linearized equations of motion follow from the perturbed second order action, the details of which are provided in \cref{app-b}. In terms of the variables defined in \eqref{h-redefn}, the action \eqref{gen-action} can be recast in a more `canonical' form as
\begin{equation}\label{eff-action1}
\begin{aligned}
S^{(2)}&=\frac{-1}{16 \pi G_N}\int dr\ d^{d}k\left[-\frac{1}{2}r^{1-d_i-z+\theta}f{H'_{xi}}^2+\frac{1}{2}r^{-1-d_i+z+\theta} {H'_{ti}}^2+k H_{ti}a'_i\right.\\
&\hspace*{2cm}+\frac{\omega^2}{2}\frac{r^{-1-d_i+z+\theta}}{f}H_{xi}^2+\frac{q^2}{2}\frac{r^{-1-d_i+z+\theta}}{f}H_{ti}^2+\frac{q\omega}{f}r^{-1-d_i+z+\theta}H_{xi}H_{ti}\\
&\hspace*{2.5cm}\left.-\frac{1}{2}fr^{d_i+3-z-\theta}{a'}^2_i+\left(\frac{\omega^2}{2}\frac{r^{d_i+1+z-\theta}}{f}-\frac{q^2}{2}r^{d_i+3-z-\theta}\right)a_i^2\right]+S^{(2)}_{bdy}\ ,
\end{aligned}
\end{equation}
which yields \eqref{nu-t-eqn}-\eqref{chi-eqn} as the equations of motion. For completion, we state the boundary action i.e. $S^{(2)}_{bdy}$ is given by
\begin{equation}
\begin{aligned}
&S^{(2)}_{bdy}=-\frac{1}{16\pi G_N}\int d^dk\ \left[2r^{1-d_i-z+\theta}fH'_{xi}H_{xi}-2r^{-1-d_i+z+\theta}H'_{ti}H_{ti}\right. \\& \left.+\frac{(\theta-d_i)f}{d_i}r^{-d_i-z+\theta}H_{xi}^2+\frac{r^{-d_i-2+\theta +z} \left(\left(d_i^2-d_i (\theta +z-4)-2 \theta \right)f-d_i(d_i-\theta +z)\right)}{2d_i f}H_{ti}^2\right]\ .
\end{aligned}
\end{equation}
Motivated by the equations of motion appearing in \eqref{nu-t-eqn}-\eqref{chi-eqn}, we observe that the coupling term between $H_{ti}$ and $a_i$ appearing in the above action \eqref{eff-action1}, namely $+kH_{ti}a'_i$  can be rewritten as $-kH'_{ti}a_i$ along with a boundary term. Thus, the effective Lagrangian reads as
\begin{equation}
\label{eq:eff-action2}
\begin{aligned}
S^{(2)}&=\frac{1}{16 \pi G_N}\int dr\ d^{d}k\left[-\frac{1}{2}r^{1-d_i-z+\theta}f{H'_{xi}}^2+\frac{1}{2}r^{-1-d_i+z+\theta} {H'_{ti}}^2-k H'_{ti}a_i\right.\\
&\hspace*{2cm}+\frac{\omega^2}{2}\frac{r^{-1-d_i+z+\theta}}{f}H_{xi}^2+\frac{q^2}{2}\frac{r^{-1-d_i+z+\theta}}{f}H_{ti}^2+\frac{q\omega}{f}r^{-1-d_i+z+\theta}H_{xi}H_{ti}\\
&\hspace*{2.5cm}\left.-\frac{1}{2}fr^{d_i+3-z-\theta}{a'}^2_i+\left(\frac{\omega^2}{2}\frac{r^{d_i+1+z-\theta}}{f}-\frac{q^2}{2}r^{d_i+3-z-\theta}\right)a_i^2\right]+S^{(2)}_{bdy}\ .
\end{aligned}
\end{equation}
The conjugate momenta for the modes $H_{xi}, H_{ti}$ and $a_i$ are defined respectively as,
\begin{equation}
\label{eq:conjugatemomenta}
16\pi G_N \Pi_{xi}=\frac{\partial \mathcal{L}}{\partial H'_{xi}}\ ,\ 16\pi G_N \Pi_{ti}=\frac{\partial \mathcal{L}}{\partial H'_{ti}}\ ,\ 16\pi G_N \Pi_{a_i}=\frac{\partial \mathcal{L}}{\partial a'_{i}}\ .
\end{equation}
The above definitions immediately yield
\begin{eqnarray}
\label{eq:Pi-xi-eqn}
&&16\pi G_N.\ \Pi_{xi}=-fr^{1-d_i-z+\theta}H'_{xi}\ ,\\
\label{Pi-ti-eqn}
&&16\pi G_N.\ \Pi_{ti}=r^{-1-d_i+z+\theta}H'_{ti}-ka_i\ ,\\
\label{eq:Pi-a-eqn}
&&16\pi G_N.\ \Pi_{a_i}=-fr^{d_i+3-z-\theta}a'_i\ .
\end{eqnarray}
Corresponding to each of the modes $H_{ti}, H_{xi}$ and $a_i$, we associate a response function given by,
\begin{equation}
\label{eq:responsefunction}
\chi(r,q,\omega)=\frac{{\Pi}(r,q,\omega)}{i\omega \phi(r,q,\omega)}\ ; \quad \phi=\{H_{xi}, H_{ti}, a_i\}\ .
\end{equation}
In terms of the response functions, the constraint equation \eqref{nu-r-eqn} takes the form 
\begin{equation}
\label{eq:constraint-eqn}
 \frac{\chi_{ti}}{\chi_{xi}}= \frac{q H_{xi}}{\omega H_{ti}}\ .
\end{equation}
Using the equations \eqref{eq:conjugatemomenta}-\eqref{eq:constraint-eqn}, we can eventually write down the generalized flow equations for the response functions $\chi_{xi}, \chi_{ti}$ and $\chi_{a_i}$ which takes the form
\begin{eqnarray}
\label{eq:chi-xifloweqn}
\partial_r \chi_{xi}&=&i\omega\left[\frac{16\pi G_N\ \chi_{xi}^2}{fr^{1-d_i-z+\theta}}-\frac{r^{z+\theta-d_i-1}}{16\pi G_N\ f}\left(1+\frac{q^2}{\omega^2}\frac{\chi_{xi}}{\chi_{ti}}\right)\right]\ ,\\
\label{eq:chi-ti-eqn}
\partial_r\chi_{ti}&=&-i\omega\left[{16\pi G_N\over r^{-1-d_i+z+\theta}}\chi_{ti}^2+{k\,a_i\over i\omega r^{-1-d_i+z+\theta }H_{ti}}\chi_{ti}+{r^{z+\theta-(d_i+1) }\over 16 \pi G_N \,f}\left({\chi_{ti}\over \chi_{xi}}+{q^2\over \omega^2}\right)\right], \nonumber \\\quad \\
\partial_r \chi_{a_i} &=& i\omega \left[ 16\pi G_N \frac{r^{z+\theta-d_i-3}}{f}\chi_{a_i}^2-\frac{r^{d_i+1+z-\theta}}{16\pi G_N f}+\frac{k^2}{16\pi G_N \omega^2}r^{d_i+1-z-\theta} \right. \nonumber\\
\label{eq:chi-aieqn}
&& \hspace*{5.5cm}\left. +{q^2 \over \omega^2}{r^{d_i+3-z-\theta} \over 16\pi G_N} - {q\ k \over \omega^3}{f r^{2-2z} \over 16\pi G_N}  {H^{\prime}_{xi} \over a_i} \right]\ .
\end{eqnarray}
Note that the above set of coupled differential equations are \emph{exact} i.e. they describe the flow of response functions for generic values of frequency and momentum. Although, they are complicated and analytically intractable, we are however interested in the hydrodynamic regime which is essentially the limit where the frequency and momenta $\omega$ and $q$ are much smaller than the temperature scale i.e. $q \ll T^{1/z} \sim r_0$ and $\omega \ll T \sim r_0^z$. Further, it is evident from \eqref{eq:chi-defn} that the $q=0$ sector is relevant for evaluating shear transport coefficient. Thus, we will focus exclusively on the $q=0$ sector of the flow equations \eqref{eq:chi-xifloweqn}-\eqref{eq:chi-aieqn}.

\section{Zero momentum response functions}\label{sec:3}

Before we proceed to study the $q \to 0$ limit of the flow equations we derived in the preceding section, it is imperative to talk about solutions of the field equations in the $q \to 0$ limit. An earlier work\cite{Mukherjee:2017ynv} analysed the field equations assuming a hydrodynamic expansion in the dimensionless parameters $\Omega=\frac{\omega}{2\pi T}$ and $Q= \frac{q}{(2\pi T)^{1/z}}$. One can however reabsorb the constant temperature factor in each term of the hydrodynamic expansion and simply write the fields as an expansion in $\omega$ and $q$. 

Starting with the equations of motion \eqref{nu-t-eqn}-\eqref{chi-eqn}, a gauge invariant combination $\cH_i$ was defined as
\begin{equation}
\label{eq:gaugeinvariantcombination}
    \cH_i=\omega H_{xi}+q H_{ti}-kq\int_{r_c}^r s^{d_i+1-z-\theta} a_i(s)\ ds
\end{equation}
The fields $\cH_i$ and $a_i$ formed a system of coupled differential equations which were solved up to first order in the hydrodynamic expansion. For the redefined field $\cH$, it was observed that for $z<d_i+2-\theta$ the terms in the hydrodynamic expansion of the field variables can be solved order-by-order. When $z=d_i+2-\theta$, the first order correction to $\cH_i$ scales logarithmically and seems to diverge close to the boundary. The logarithmic scaling is suggestive of the emergence of a new scale in the UV limit. In the regime when $z>d_i+2-\theta$, the first order correction to $\cH_i$ diverges suggesting a breakdown of the methodology for parameters in this regime. The solution to the combination $\cH_i$ up to first order in the hydrodynamic expansion is given by
\begin{equation}
\label{eq:earlierHisoln}
    \cH_i= C_0 f(r)^{-\frac{i\Omega}{2}}\left[1+\frac{iq^2}{(d_i+2-z-\theta)\omega}r_0^{z-2}\cdot (1-(r_0 r)^{d_i+2-z-\theta}) \right]\ .
\end{equation}
The gauge field fluctuations, $a_i$ satisfy a second order non-homogeneous differential equation. Upon imposing regularity on $a_i$, the leading solution takes the form
\begin{equation}
	\label{eq:aiqneq0-soln}
	a_i=-iC_0k \frac{q}{\omega}\cdot \frac{r_0^{d_i-\theta}}{(d_i+z-\theta)^2}f(r)^{1-\frac{i\omega}{4\pi T}}(r_0r)^{-(d_i+z-\theta)}\ ,
\end{equation}
where $C_0$ is an arbitrary non-zero constant. The first order piece does not have a closed form solution but can be written as an integral and thus cannot give us further insight into its behaviour. The reader can find details and methodology of solving for $a_i$ up to first order in \cite{Mukherjee:2017ynv}. Since in the current context, our interest is to explore the flow equations, we will not further concern ourselves with solutions to the fields $H_{xi}, H_{ti}$. We will however make certain assumptions about them which will help us in dealing with the complicated flow equations we derived in the preceding section.

Motivated from the form of $\cH_i$ and $a_i$ as given in \eqref{eq:earlierHisoln} and \eqref{eq:aiqneq0-soln}, we will assume that the perturbations $H_{xi}, H_{ti}$ and $a_i$ for $\omega \neq 0$ and $q \neq 0$ admit a hydrodynamic expansion of the form 
\begin{equation}
\label{eq:fieldhydroexpnsn}
    \phi(r, \omega,q)= \phi^{(-1)}(r,\omega,q)+\phi^{(0)}(r,\omega,q)+\phi^{(1)}(r,\omega,q)+\cdots=\sum_{n=-1}^{\infty}\phi^{(n)}(r,\omega,q)\ ,
\end{equation}
where $\phi(r,\omega,q)$ represents any one of the perturbative modes $H_{xi}, H_{ti}$ or $a_i$. The leading term $\phi^{(-1)}(r,\omega,q)$ is parametrically an $\mathcal{O}(\frac{1}{\omega})$ quantity while $\phi^{(n)}(r,\omega,q) \sim \mathcal{O}(\omega^n) \sim \mathcal{O}(q^n)$. The first term in the above expression can be generically of the form
\begin{equation}
\label{eq:pole-term}
    \phi^{(-1)}(r,\omega,q) \sim \sum_{a \geq 0} \frac{q^a}{\omega^{a+1}}\cb_{a}(r)
\end{equation}
while the $\mathcal{O}(1)$ term and the $n$-th order term in the hydrodynamic expansion will take the general schematic form
\begin{equation}
\label{eq:schematic-nth-order}
    \phi^{(0)}(r,\omega,q) \sim \sum_{a \geq 0}\frac{q^a}{\omega^a}\cg_{a}(r)\ \text{and}\ \phi^{(n)}(r,\omega,q) \sim \mathop{\sum_{a,b}}_{k>0}q^a\omega^b\frac{q^k}{\omega^k}\ch_{(a,b,k)}(r)\ ,
\end{equation}
respectively. In \eqref{eq:pole-term} and \eqref{eq:schematic-nth-order}, all the exponents $a,b$ and $k$ are strictly positive. The functions $\cb_a(r), \cg_a(r)$ and $\ch_{(a,b,k)}(r)$ are all regular in the interval $0<r<\frac{1}{r_0}$. The sum in the $n$-th order term has a `prime' to denote that it is a \emph{constrained} sum such that $a+b=n \geq 1$. The above schematic forms of each term in the hydrodynamic expansion of the field variables is well behaved in the limit $q \to 0$.

A comparison of \eqref{eq:aiqneq0-soln} with the schematic forms as given in \eqref{eq:pole-term} and \eqref{eq:schematic-nth-order} tells us that for $a_i$, the $\mathcal{O}(\frac{1}{\omega})$ term is identically zero; the $\mathcal{O}(1)$ term consists of a single term with $a=1$ while $\cg_1(r) \sim f(r)^{-\frac{i\Omega}{2}}$. A comparison of the above schematic expansion with $\cH_i$ as given in \eqref{eq:earlierHisoln} is difficult, since it appears as as a linear combination of $H_{xi},H_{ti}$ and an integral over $a_i$. We can still comment on the heuristic behaviour of the response functions that follows from the above assumptions regarding the hydrodynamic expansion of the field variables.

The structure of \eqref{eq:Pi-xi-eqn}-\eqref{eq:responsefunction} along with \eqref{eq:pole-term}-\eqref{eq:schematic-nth-order} implies that the response function can be written schematically as
\begin{equation}
    \chi \simeq \frac{F(r)}{i\omega}\left(\chi^{(0)}(r,\omega,q)+\chi^{(1)}(r,\omega,q)+\chi^{(2)}(r,\omega,q)+\cdots \right)
\end{equation}
where $\chi^{(n)}(r,\omega,q)$ denotes a term which is $\mathcal{O}(q^n)\sim \mathcal{O}(\omega^n)$ in the hydrodynamic expansion but is determined by the explicit forms of $\cb_a(r), \cg_a(r)$ and $\ch_{(a,b,k)}(r)$ while $F(r)$ is some specific function depending on which mode is under consideration. More specifically,
\begin{equation}
\chi^{(0)}(r,\omega,q)=\frac{\sum_{a \geq 0} \frac{q^a}{\omega^{a+1}}\partial_r \cb_{a}(r)}{\sum_{a \geq 0} \frac{q^a}{\omega^{a+1}}\cb_{a}(r)}\ , \chi^{(1)}(r,\omega,q)=\frac{\sum_{a,b \geq 0}\frac{q^{a+b}}{\omega^{a+b+1}}\mathcal{W}[\cb_a(r),\cg_b(r)]}{\sum_{a,b \geq 0}\frac{q^{a+b}}{\omega^{a+b+2}}\cb_a(r)\cb_b(r)}
\end{equation}
where $\mathcal{W}[f,g]=fg'-f'g$ denotes the Wronskian for the pair of functions $f$ and $g$. In the case, neither of these are linear combinations of various powers of the ratio $\frac{q}{\omega}$, we simply recover 
\begin{equation}
\begin{aligned}
    \chi^{(0)}(r,\omega,q)=\partial_r \ln \cb_a (r)\ \ &,\ \ \chi^{(1)}(r,\omega,q)=\omega \frac{\mathcal{W}[\cb_a(r),\cg_b(r)]}{\cb_a(r)\cb_b(r)}\ .
\end{aligned}
\end{equation}
Armed with the above heuristic analysis, we further closely look at the following terms appearing in \eqref{eq:chi-xifloweqn}-\eqref{eq:chi-aieqn}.
\begin{itemize}
    \item The last term appearing in \eqref{eq:chi-xifloweqn} can be written as
    \begin{equation}
        H(r)\frac{q^2}{\omega}\frac{\chi_{xi}}{\chi_{ti}},
    \end{equation}
    where $H(r)$ is a function of $r$ whose details we are not concerned with for the purpose of this analysis. For the sake of simplicity, if we assume \eqref{eq:pole-term} is not in fact a linear combination of various powers of the ratio $\frac{q}{\omega}$, the leading behavior of this term \emph{for non-zero $q$ and $\omega$} is given by
    \begin{equation}
        H(r)\frac{q^2}{\omega}\frac{\chi_{xi}}{\chi_{ti}} \simeq \frac{q^2}{\omega}\left(\tilde{H}_1(r)+\omega \tilde{H}_2(r)+\cdots \right),
    \end{equation}
    where the ``$\cdots$" represents terms that are higher order in $q$ or $\omega$ while $\tilde{H}_n(r)$ represents various functions of $r$. Now, every expression $\tilde{H}_n(r)$ involve ratios of derivatives of the family of functions $\cb_a(r), \cg_a(r)$ and $\ch_{(a,b,k)}(r)$ that appear in the hydrodynamic expansion of the field variables. To be more explicit, 
    \begin{equation}
        \tilde{H}_1(r) \sim \frac{\partial_r \cb_a^{(xi)}(r)}{\partial_r \cb_a^{(ti)}(r)}.
    \end{equation}
    At this point, we further make the assumption that for non-zero $q$ and $\omega$, each of these functions i.e. $\cb_a(r), \cg_a(r)$ and $\ch_{(a,b,k)}(r)$ appearing in the field expansion of $H_{ti}$ and $H_{xi}$ are \emph{non-constant, non-trivial functions of $r$}. Clearly, under the above assumption, this term vanishes when $q \to 0$. We will subsequently infer from the equations of motion at $q=0$ that $\chi_{ti}=0$ for this sector however, the term that we just discussed does not have any singularity or does not go to any constant as $q \to 0$. Physically speaking, $\chi_{ti}$ presumably contains a  leading $O(q)$ piece which ensures that the ratio $\frac{q^2}{\chi_{ti}} \to 0$ as $q \to 0$ while presence of higher powers of the momenta $q$ in subsequent higher order terms ensure it vanishes identically as $q \to 0$.
    
    \item In \eqref{eq:chi-ti-eqn}, we see the last two terms can be schematically written as
    \begin{equation}
        \omega G(r) \left(\frac{\chi_{ti}}{\chi_{xi}}+\frac{q^2}{\omega^2} \right) \xrightarrow{q \to 0} \omega G(r)\frac{\chi_{ti}}{\chi_{xi}}\ .
    \end{equation}
    Although this term will indeed vanish at $q=0$, since $\chi_{ti}=0$ in this sector, our assumptions up to this point dictates a possible $O(\omega)$ contribution as to the flow equations as $q \to 0$. Hence, we keep this term in the limit of vanishing momenta. 
    
    \item Finally, we need to study the final term in \eqref{eq:chi-aieqn} which we schematically write as
    \begin{equation}
        \frac{q}{\omega^2}P(r)\frac{H'_{xi}}{a_i}.
    \end{equation}
    Note that from \eqref{eq:aiqneq0-soln} and comparing with the expansion \eqref{eq:fieldhydroexpnsn}, it follows that for the field $a_i$, the family of functions $\cb_a(r)=0$ identically. Again, for simplicity, assuming $H_{xi}$ does not have have a linear combination of terms at $\mathcal{O}(1/\omega)$, we get the leading behaviour of the last term as
    \begin{equation}
        \frac{q}{\omega^2}P(r)\frac{H'_{xi}}{a_i} \simeq \frac{q^a}{\omega^{a+2}}P(r)\frac{\partial_r \cb_a^{(xi)}(r)}{g_1^{(a_i)}(r)} \xrightarrow{q \to 0} 0\ .
    \end{equation}
\end{itemize}
Thus, in the $q \to 0$ limit, we recover the simplified flow equations as
\begin{eqnarray}
\label{eq:zeroqfloweqn1}
\partial_r \chi_{xi}&=&i\omega\left[\frac{16\pi G_N\ \chi_{xi}^2}{fr^{1-d_i-z+\theta}}-\frac{r^{z+\theta-d_i-1}}{16\pi G_N\ f}\right]\ ,\\
\label{eq:zeroqfloweqn2}
\partial_r\chi_{ti}&=&-i\omega\left[{16\pi G_N\over r^{-1-d_i+z+\theta}}\chi_{ti}^2+{k\,a_i\over i\omega r^{-1-d_i+z+\theta }H_{ti}}\chi_{ti}+{r^{z+\theta-(d_i+1) }\over 16 \pi G_N \,f}{\chi_{ti}\over \chi_{xi}}\right], \nonumber \\\quad \\
\label{eq:zeroqfloweqn3}
\partial_r \chi_{a_i} &=& i\omega \left[ 16\pi G_N \frac{r^{z+\theta-d_i-3}}{f}\chi_{a_i}^2-\frac{r^{d_i+1+z-\theta}}{16\pi G_N f}+\frac{k^2}{16\pi G_N \omega^2}r^{d_i+1-z-\theta} \right]\ .
\end{eqnarray}
The above equations can also be derived by turning on perturbations of the form $e^{-i\omega t}h_{\mu \nu}(r)$ and choosing the radial gauge $h_{\mu r}=0$. Before we proceed with the detailed analysis of the flow of response functions, the equations of motion in the $q=0$ sector simplifies significantly to give
\begin{eqnarray}
\label{eq:qzeroeq1}
\partial_r(r^{z+\theta-(d_i+1)}H'_{ti}-ka_i) &=&0\ ,\\
\label{eq:qzeroeq2}
\partial_r(r^{\theta-z-d_i+1}fH'_{xi})+\frac{r^{z+\theta-(d_i+1)}}{f}\ \omega^2  H_{xi}&=&0\ ,\\
\label{eq:qzeroeq3}
H'_{ti}-kr^{(d_i+1)-z-\theta}a_i&=&0\ ,\\
\label{eq:qzeroeq4}
\partial_r(r^{d_i+3-z-\theta}f a'_i)+\frac{r^{d_i+1+z-\theta}}{f}\omega^2 a_i -kH'_{ti}&=&0\ .
\end{eqnarray}
Thus, in the $q=0$ sector, the mode $H_{xi}$ further decouples from $H_{ti}$ and $a_i$. The constraint equation \eqref{eq:qzeroeq3} clearly implies
\begin{equation}\label{eq:piti}
   \left. \Pi_{ti}\right|_{q=0}=0\ .
\end{equation}
By the assumptions we made in \eqref{eq:fieldhydroexpnsn}-\eqref{eq:schematic-nth-order}, we see that
\begin{equation}
    \lim_{q \to 0}\chi_{ti}=\lim_{q \to 0}\frac{\Pi_{ti}}{i\omega H_{ti}}=0\ .
\end{equation}
This indeed is consistent with \eqref{eq:zeroqfloweqn2} and renders the equation trivial. Thus the $q=0$ sector requires us to analyse two \emph{independent} equations governing the flow of $H_{xi}$ and $a_i$ given by \eqref{eq:zeroqfloweqn1} and \eqref{eq:zeroqfloweqn3} respectively. 

\subsection{Response function $\chi_{xi}$ at $q=0$} \label{sec:chixi}
As argued in the previous section, in the $q=0$ sector, the $\chi_{xi}$ flow equation decouples from the $\chi_{ti}$ flow equation and \eqref{eq:chi-xifloweqn} simplifies to
\begin{equation}
\label{eq:chi-xiqzeroeqn}
\partial_r \chi_{xi}=\frac{i\omega}{f}\left[\frac{16\pi G_N}{r^{1-d_i-z+\theta}}\chi _{xi}^2-\frac{r^{z+\theta-d_i-1}}{16\pi G_N}\right]\ .
\end{equation}
If we demand regularity of $\chi_{xi}$ at the horizon, we clearly see the RHS of the above is singular at $r =\displaystyle{\frac{1}{r_0}}$. This forces us to choose
\begin{equation}
    \left[\frac{16\pi G_N}{r^{1-d_i-z+\theta}}\chi _{xi}^2-\frac{r^{z+\theta-d_i-1}}{16\pi G_N}\right]_{r=\frac{1}{r_0}}=0\ ,
\end{equation}
leading to the boundary condition  
\begin{equation}
\label{eq:chixi-boundarycond}
    \chi_{xi}\left(\frac{1}{r_0},\omega\right)=\frac{r_0^{d_i-\theta}}{16\pi G_N}.
\end{equation}
In the hydrodynamic regime, we are allowed to write a perturbative expansion for $\chi_{xi}(r,\omega)$ as
\begin{equation}
\chi_{xi}(r,\omega)=\chi_{xi}^{(0)}(r)+\omega \chi_{xi}^{(1)}(r)+\mO(\omega^2) ,
\end{equation}
where $\mO(\omega^2)$ represents higher order terms beyond the linear one. Plugging in the above expansion, in \eqref{eq:chi-xiqzeroeqn}, the leading order piece follows
\begin{equation}
\label{chi-leading}
\partial_r \chi_{xi}^{(0)}(r)=0\ .
\end{equation}
Physically, the above equation tells us that the RG flow of the $\chi_{xi}$ is trivial at leading order remaining unchanged as we go along the radial direction. Along with the boundary condition \eqref{eq:chixi-boundarycond} that we just derived, we have
\begin{equation}
    \chi_{xi}^{(0)}(r)=\frac{r_0^{d_i-\theta}}{16\pi G_N}\ .
\end{equation}
The $\mO(\omega)$ equation which gives the flow of $\chi_{xi}^{(1)}$, is given by,
\begin{equation}
\partial_r \chi_{xi}^{(1)}=-i\frac{r^{-d_i-1+z+\theta}}{16\pi G_N \ f(r)}\left(1-(r_0r)^{2(d_i-\theta)}\right)\ .
\end{equation}
The solution to the above equation is
\begin{equation}
\label{chixi-firstorder}
\begin{aligned}
\chi^{(1)}_{xi}(r)&=-\frac{ir_0^{d_i-z-\theta}}{16\pi G_N}\left[\frac{(r_0r)^{-d_i+z+\theta}}{z+\theta-d_i}\ _2F_1\Big[1,1-\frac{2(d_i-\theta)}{d_i+z-\theta},2-\frac{2(d_i-\theta)}{d_i+z-\theta};(r_0r)^{d_i+z-\theta}\Big]\right.\\
&\hspace*{7cm}\left. +\frac{\log f(r)}{d_i+z-\theta}+C\right]\ \mbox{when}\ z \neq d_i-\theta\ ,\\
&=-\frac{i}{16\pi G_N}\log \frac{r}{C'}\ \mbox{when}\ z=d_i-\theta\ ,
\end{aligned}
\end{equation}
where $C$ and $C'$ are integration constants for the two cases of the Lifshitz exponent $z$ while $_2F_1[a,b,c;r]$ represents the hypergeometric function. We then come across the following two cases,

\paragraph{\underline{Case I $\bullet$ $z \neq d_i-\theta$:}}
Using the boundary condition \eqref{eq:chixi-boundarycond}, we can fix the constant of integration to be
\begin{equation}
C=\frac{(\gamma+\psi(\frac{z-d_i+\theta}{d_i+z-\theta}))}{d_i+z-\theta}
\end{equation}
where $\gamma$ is the Euler-Mascheroni constant and $\psi(x)$ is the polygamma function which is singular over the set non-positive definite integers. Taking into account the null energy condition \eqref{nullee}, we focus when $d_i-\theta>0$ and $z>1$. Since, this solution is true when $z \neq d_i-\theta$, the argument in the polygamma function cannot be 0. However, $\frac{z-d_i+\theta}{d_i+z-\theta}=-1$ gives $z=0$ which violates our the assumption of $z\geq 1$. For all other parameter values of $(z,\theta)$ the null energy condition ensures that $\psi(\frac{z-d_i+\theta}{d_i+z-\theta})$ is non-singular.

\paragraph{\underline{Case II $\bullet$ $z = d_i-\theta$:}} In this case too, plugging in the boundary condition \eqref{eq:chixi-boundarycond}, we get,
\begin{equation}
C'=\frac{1}{r_0}
\end{equation}
which then gives the full solution
\begin{equation}
\chi_{xi}(r)=\frac{r_0^{d_i-1}}{16 \pi G_N}-\frac{i\omega}{16\pi G_N}\log(r_0r)\ .
\end{equation}
Clearly the divergent nature of the solutions as $r \rightarrow 0$, hints at a possible breakdown of the analysis when $z=d_i-\theta$ near the boundary. 

Earlier works \cite{Kolekar:2016yzg,Mukherjee:2017ynv} used perturbative techniques to evaluate 2-point correlator of the stress-energy tensor which seemingly broke down when $z>d_i+2-\theta$. However, an analysis of the response function corresponding to $H_{xi}$ i.e. $\chi_{xi}$ seems to carry through for all values of the Lifshitz exponent. As mentioned earlier in \eqref{eq:chi-defn}, shear viscosity up to leading order is thus given by 
\begin{equation}
	\eta=\chi_{xi}=
		\frac{r_0^{d_i-\theta}}{16 \pi G_N}
\end{equation}
which inturn saturates the KSS bound of $\frac{\eta}{s}=\frac{1}{4\pi}$. Also, note that the first order correction for either cases, namely $z=d_i-\theta$ and $z \neq d_i-\theta$ is positive since $r_0r<1$ thus following the bound. However, when $z=d_i-\theta$, we see the first order correction to be logarithmic and is actually divergent at the boundary  when $r \to 0$. This enforces us to put a cut-off suggesting the emergence of a new scale.

Interestingly, earlier works\cite{Balasubramanian:2010uk,Donos:2010tu} constructed families of Lifshitz geometries as dimensional reduction of $AdS$ null deformations. Specifically, starting with $AdS_5$ null deformation, one can perform a reduction along one of the light-cone coordinates, namely $x^{+}$ which results in a 4-dimensional metric of the form \eqref{hvmetric} with $z=d_i=2$ and $\theta=0$. Thus, dimensional reduction of null deformed $AdS_5$ results in a metric which falls in the family of hvLif solutions constrained by $z=d_i-\theta$. In light of this observation, it will be interesting to understand the logarithmic scaling of the first order contribution to $\chi_{xi}$ from the perspective of the deformed higher dimensional theory.

\subsection{Response function $\chi_{a_i}$ at $q=0$} \label{sec:chiti}

Recall from our earlier definition \eqref{eq:responsefunction}, that the response function $\chi_{a_i}$ associated to $a_i$ is defined as
\begin{equation}
    \chi_{a_i}=\frac{\Pi_{a_i}}{i\omega a_i}
\end{equation}
To reiterate, the flow equation for the response function $\chi_{a_i}$ decouples from that of $\chi_{xi}$ and $\chi_{ti}$ in the limit $q \rightarrow 0$ to yield,
\begin{equation}
\label{eq:chi-a-eqn}
    \partial_r \chi_{a_i}=i\omega \left(16\pi G_N \frac{r^{z+\theta-d_i-3}}{f}\chi_{a_i}^2-\frac{r^{d_i+1+z-\theta}}{16\pi G_N f}+\frac{k^2}{16\pi G_N \omega^2}r^{d_i+1-z-\theta} \right)\ .
\end{equation}
The structure of \eqref{eq:chi-a-eqn} is significantly different from the flow equation of $\chi_{xi}$. Assuming a Laurent expansion in $\omega$ for the function $\chi_{a_i}(r,\omega)$, we see that in general it must have a term which goes as $\frac{1}{\omega}$ along with regular terms. Thus, like the earlier case of $\chi_{xi}$, it does not make sense to naively perform a hydrodynamic expansion of containing only positive powers of $\omega$. However, we define the new field
\begin{equation}
\label{eq:zeta-defn}
    \zeta_{a_i}=\omega \chi_{a_i}\ ,
\end{equation}
in terms of which \eqref{eq:chi-a-eqn} becomes
\begin{equation}
\label{eq:zeta-eqn}
    \partial_r \zeta_{a_i}(r,\omega)= i \left[16 \pi G_N \frac{r^{z+\theta-d_i-3}}{f}\zeta_{a_i}^2(r,\omega) +\frac{k^2 r^{d_i-z-\theta+1}}{16 \pi G_N}-\omega^2 \frac{r^{d_i+z-\theta+1}}{16 \pi G_N f}\right]
\end{equation}
Imposing regularity for $\zeta_{a_i}$ along the radial direction demands us to write the boundary condition as
\begin{equation}
\left[ 16\pi G_N r^{z+\theta-d_i-3}\zeta_{a_i}^2\left(r,\omega\right)-\frac{\omega^2 r^{d_i+1+z-\theta}}{16\pi G_N }\right]_{r= \frac{1}{r_0}}=0\ . 
\end{equation}
which eventually yields,
\begin{equation}
\label{eq:chi-a-boundarycond}
    \zeta_{a_i}\left(\frac{1}{r_0},\omega\right)=\omega\frac{r_0^{\theta-d_i-2}}{16 \pi G_N}\ .
\end{equation}
One must note that \eqref{eq:zeta-eqn} is exact in $\omega$ and consistent with a hydrodynamic expansion of the form
\begin{equation}
    \zeta_{a_i}(r,\omega)=\zeta_{a_i}^{(0)}(r)+\omega \zeta_{a_i}^{(1)}(r)+ \omega^2 \zeta_{a_i}^{(2)}(r)+\cdots\ .
\end{equation}
Also, the demand of regularity gives us the $\zeta_{a_i}$ at the horizon which depends explicitly on the frequency $\omega$. Thus, regularity in the context of the above hydrodynamic expansion implies $\zeta^{(m)}_{a_i}(1/r_0)=0$ for all $m \neq 1$ while $\zeta^{(1)}_{a_i}(1/r_0)= \frac{r_0^{\theta-d_i-2}}{16 \pi G_N}$. Unlike the earlier case of $\chi_{xi}$, we see here that at leading order $\partial_r \zeta_{a_i}$ follows a nontrivial flow equation given by
\begin{equation}
    \partial_r \zeta_{a_i}^{(0)}(r)= i \left[16 \pi G_N \frac{r^{z+\theta-d_i-3}}{f}{\zeta_{a_i}^{(0)}(r)}^2 +\frac{k^2 r^{d_i-z-\theta+1}}{16 \pi G_N}\right]\ .
\end{equation}

Thus, we see for this response function, the RG flow is not trivial and it actually changes along the radial direction. Solving the above equation yields complicated solutions which one cannot use easily to construct further subleading contributions that are higher order in $\omega$.

To circumvent the issue, we follow a different strategy. We will analyse the flow equation successively in the \emph{near horizon} and the \emph{near boundary} region.

\paragraph{\underline{Near horizon region:}} In order to analyse the flow near the horizon, we define a new radial coordinate $\rho$ given by
\begin{equation}
\label{eq:nearbdy-radial}
    \rho={1\over r_0}-r\ .
\end{equation}
In turn, the blackening factor can be written as
\begin{equation}
    f=(d_i+z-\theta)r_0\rho +\mathcal{O}(\rho^2)\ .
\end{equation}
Thus in the near horizon region, the flow equation can be approximated as
\begin{equation}
\label{eq:zetanearhorizon}
\begin{aligned}
    \partial_{\rho}\nhzeta(\rho, \omega)&=-i\left[16 \pi G_N \frac{r_0^{d_i+2-z-\theta}}{d_i+z-\theta}\frac{(1+(d_i+3-z-\theta)r_0 \rho)}{\rho}\nhzeta(\rho, \omega)^2 \right.\\
    &\hspace*{3.2cm}\left. -\frac{\omega^2}{16 \pi G_N}\frac{r_0^{-d_i-z+\theta-2}}{d_i+z-\theta}\frac{\left(1-(d_i+z-\theta+1)r_0 \rho \right)}{\rho} \right]\ .
\end{aligned}
\end{equation}
An ansatz consistent with a hydrodynamics description may be written as
\begin{equation}
    \nhzeta(\rho, \omega)=\nhzeta^{(0)}(\rho)+\omega \nhzeta^{(1)}(\rho)+ \omega^2 \nhzeta^{(2)}+ \mathcal{O}(\omega^3)\ .
\end{equation}
It is clear from \eqref{eq:zetanearhorizon} that the second term on the RHS affects only at $\mathcal{O}(\omega^2)$. Also, the boundary condition \eqref{eq:chi-a-boundarycond} implies that $\nhzeta^{(0)}(0)=\nhzeta^{(2)}(0)=0$. The resulting equation for $\nhzeta^{(0)}(\rho)$ is given by,
\begin{equation}
    \partial_{\rho}\nhzeta^{(0)}(\rho)=-16 i \pi G_N \frac{r_0^{d_i+2-z-\theta}}{d_i+z-\theta}\frac{(1+(d_i+3-z-\theta)r_0 \rho)}{\rho}\nhzeta^{(0)}(\rho)^2.
\end{equation}
Which on solving naively yields a solution of the form $-\frac{i}{c_1+\mathcal{A}\rho +\mathcal{B}\log \rho}$ where $\mathcal{A}$ and $\mathcal{B}$ are constants depending on $r_0, d_i, z$ and $\theta$ while $c_1$ is an arbitrary constant which remains unfixed even after imposing the relevant boundary condition for $\nhzeta^{(0)}(\rho)$. This is because the very boundary condition \eqref{eq:chi-a-boundarycond} is specified at a singular point of the equation. We can however choose a cutoff surface at $\rho=\epsilon$ (which can be thought of as a \emph{stretched membrane}) hovering at a distance $\epsilon$ outside the real horizon at $\frac{1}{r_0}$ where $\nhzeta^{(0)}(\epsilon)=0$ which then implies 
\begin{equation}
    \nhzeta^{(0)}(\rho)=0\ ,
\end{equation}
identically in the \emph{near horizon region}. This in turn leads to the simple equation at $\mathcal{O}(\omega)$ i.e.
\begin{equation}
    \partial_{\rho} \nhzeta^{(1)}(\rho)=0\ .
\end{equation}
The above along with \eqref{eq:chi-a-boundarycond} implies
\begin{equation}
    \nhzeta^{(1)}(\rho)=\frac{r_0^{\theta-d_i-2}}{16 \pi G_N}\ .
\end{equation}
Eventually, the equation at $\mathcal{O}(\omega^2)$ is given by
\begin{equation}
    \partial_{\rho}\nhzeta^{(2)}(\rho)= -\frac{i(d_i+2-\theta)}{8\pi G_N (d_i+z-\theta)}r_0^{-d_i-z+\theta-1}\ .
\end{equation}
The solution to the above equation consistent with the boundary condition \eqref{eq:chi-a-boundarycond} is
\begin{equation}
    \nhzeta^{(2)}(\rho)=-\frac{i(d_i+2-\theta)}{8\pi G_N (d_i+z-\theta)}r_0^{-d_i-z+\theta-1} \rho\ .
\end{equation}
Thus in the \emph{near horizon region}, we have,
\begin{equation}
    \nhzeta(\rho, \omega) \approx \omega \frac{r_0^{\theta-d_i-2}}{16 \pi G_N} -\frac{i\omega^2(d_i+2-\theta)}{8\pi G_N (d_i+z-\theta)}r_0^{-d_i-z+\theta-1} \rho\ . 
\end{equation}
Using \eqref{eq:zeta-defn} and \eqref{eq:nearbdy-radial}, we see that in the \emph{near horizon region}, we can write, 
\begin{equation}
    \chi_{a_i}(r,\omega) \approx \frac{r_0^{\theta-d_i-2}}{16 \pi G_N} -\frac{i\omega(d_i+2-\theta)}{8\pi G_N (d_i+z-\theta)}r_0^{-d_i-z+\theta-1} \left(\frac{1}{r_0}-r \right)\ .
\end{equation}
Clearly, $\chi_{a_i}$ being a constant at leading order exhibits trivial RG flow and is thus qualitatively similar to $\chi_{xi}$. However, one must note that this is true only in the near horizon region.

\paragraph{\underline{Near boundary region:}} In this regime, we can approximate the blackening factor $f(r) \approx 1$ which simplifies \eqref{eq:zeta-eqn} to 
\begin{equation}
\label{eq:nearbdyzeta}
    \partial_r \zeta_{a_i}(r,\omega)= i \left[16 \pi G_N r^{z+\theta-d_i-3}\zeta_{a_i}^2(r,\omega) +\frac{k^2 r^{d_i-z-\theta+1}}{16 \pi G_N}-\omega^2 \frac{r^{d_i+z-\theta+1}}{16 \pi G_N}\right]\ .
\end{equation}
Assuming a series expansion in $\omega$ of the form
\begin{equation}
    \zeta_{a_i}(r, \omega) \xrightarrow{r \rightarrow 0} \bzeta^{(0)}(r)+\omega \bzeta^{(1)}(r)+\omega^2 \bzeta^{(2)}(r)+\mathcal{O}(\omega^3)\ ,
\end{equation}
we see that the leading order satisfies an equation of the form
\begin{equation}
    \partial_r \bzeta^{(0)}(r)= i \left[16 \pi G_N r^{z+\theta-d_i-3}\bzeta^{(0)}(r)^2 +\frac{k^2 r^{d_i-z-\theta+1}}{16 \pi G_N}\right]\ ,
\end{equation}
whose solution is given by
\begin{equation}
\label{eq:bzetaleading}
    \bzeta^{(0)}(r)= \frac{ir^{d_i+2-z-\theta}}{32 \pi G_N}\left[(d_i+3z-\theta-2)\frac{c_1-r^{-(d_i+3z-\theta-2)}}{c_1+r^{-(d_i+3z-\theta-2)}}-(d_i+2-z-\theta) \right]\ .
\end{equation}
Assuming reality of the gauge field \eqref{bg-gauge-fld} i.e. $z>1$ the null energy condition \eqref{nullee} implies $d_i+z-\theta>0$ which in turn implies $d_i+3z-\theta-2=(d_i+z-\theta)+2(z-1) >0$. Thus, near the boundary, 
\begin{equation}
    \lim_{r \to 0} (d_i+3z-\theta-2)\frac{c_1-r^{-(d_i+3z-\theta-2)}}{c_1+r^{-(d_i+3z-\theta-2)}}-(d_i+2-z-\theta) =-2(d_i+z-\theta)\ .
\end{equation} 
Clearly, when $d_i+2-z-\theta>0$, $\bzeta^{(0)}(r) \rightarrow 0$ as $r \rightarrow 0$, however for $d_i+2-z-\theta <0$, we see a divergent solution as $r \rightarrow 0$ while it goes to a constant as $r \to 0$ when $z=d_i+2-\theta$. In fact, due to the functional form of the solution \eqref{eq:bzetaleading}, its limit as $r \rightarrow 0$ will be independent of the constant $c_1$ which will remain unfixed for any Dirichlet condition imposed at the boundary. Hence, for $z>d_i+2-\theta$ it seems such a \emph{hydrodynamic} description for the gauge field response function will simply breakdown near the boundary. 

Starting with a $AdS_{d_i+3}$ dimensional boosted black brane, performing a boost and taking an appropriate double scaling limit involving the boost parameter and horizon radius yields the so-called $AdS_{d_i+3}$ plane wave. Subsequently reducing along $x^{+}$ and identifying $x^{-}\equiv t$ yields \eqref{hvmetric} where the Lifshitz exponent $z$ and hyperscaling violating exponent $\theta$ are related by\cite{Narayan:2012hk}
\begin{equation}
    z=\frac{d_i+4}{2} \quad \text{and} \quad \theta=\frac{d_i}{2}\ .
\end{equation}
Clearly, from the above expressions it follows that $z=d_i+2-\theta$. This is precisely the point in the $(z,\theta)$ parameter space where we see the leading behaviour of $\zeta_{a_i}$ near the boundary is a constant. From the viewpoint of the $AdS_{d_i+3}$ boosted black brane, this is suggestive that the hydrodynamic analysis for such effective theories obtained as null reductions break down. However, a concrete understanding of this breakdown would require further detailed analysis concerning the stability of such spacetimes which we plan to carry out in subsequent works.     

It is interesting to notice that this condition was recovered in earlier works \cite{Kolekar:2016yzg, Mukherjee:2017ynv}. In particular, \cite{Mukherjee:2017ynv} studied QNM modes in the black brane background given by \eqref{hvmetric}. As described earlier in section \ref{sec:3}, the gauge invariant combination \eqref{eq:gaugeinvariantcombination} has a solution given by \eqref{eq:earlierHisoln} up to first order in the hydrodynamic expansion for $z< d_i+2-\theta$. For $z=d_i+2-\theta$, the first order term develops a logarithmic scaling while it diverges near the boundary when $z>d_i+2-\theta$. Our current analysis suggests it is the behaviour of the perturbations in the background gauge field i.e. $a_i$ near the boundary which is presumably the cause of this divergence. Thus, the RG analysis seems to be suggestive of the fact that it is the hydrodynamic expansion of $a_i$ which breaks down causing its response function to yield an \emph{unphysical} answer when $z>d_i+2-\theta$. Further, we should contrast this with \cite{Kolekar:2016pnr,Kolekar:2016yzg} which were near-horizon analysis, also led to the same restriction on the Lifshitz exponent $z$. In our current analysis, the divergence seem to occur in the boundary theory as $r \to 0$. Earlier work \cite{Ghodrati:2014spa} studied hvLif solutions as solutions to theories with higher derivative corrections. Null energy conditions and stability criteria led to certain regions in the $(z,\theta)$ parameter space that were identified as \emph{physically allowed}. The criteria that we obtain above i.e. $z<d_i+2-\theta$ seems to be an independent bound which cannot be obtained by NECs or stability criteria.     

The first order equation is given by
\begin{equation}
     \partial_r \bzeta^{(1)}(r) =32i \pi G_N r^{z+\theta-d_i-3}\bzeta^{(0)}(r)\bzeta^{(1)}(r)\ ,
\end{equation}
which has a solution of the form
\begin{equation}
    \bzeta^{(1)}(r)=c_2\frac{r^{2(d_i+z-\theta)}}{(1+c_1 r^{d_i+3z-\theta-2})^2}\ .
\end{equation}
Owing to the null energy condition \eqref{nullee} and reality of the gauge fields, which implies $z>1$, we see that 
\begin{equation}
    \lim_{r \to 0} \bzeta^{(1)}(r) =0\ ,
\end{equation}
which leaves the constant $c_2$ which remains unfixed. Finally, the equation governing the second order contribution is given by
\begin{equation}
\label{eq:zetabdy2eqn}
\begin{aligned}
\partial_r \bzeta^{(2)}(r)&+ \frac{2[2(z-1)c_1r^{d_i+3z-\theta-2}-(d_i+z-\theta)]}{r(c_1 r^{d_i+3z-\theta-2}+1)} \bzeta^{(2)}(r)\\
&\hspace*{1.5cm}+i \left(\frac{r^{2z-1}}{16 \pi G_N}-16 \pi G_N c_2^2 \frac{r^{3d_i+5z-3\theta-3}}{(c_1 r^{d_i+3z-\theta-2}+1)^4}\right)=0\ ,
\end{aligned}    
\end{equation}
whose solutions are listed in \cref{app:zetaabdy-soln}. From \eqref{eq:zeta-defn}, it follows that the response function associated to $a_i$ near the boundary is given by
\begin{equation}
\begin{aligned}
    \chi_{a_i}(r, \omega) &\approx \frac{ir^{d_i+2-z-\theta}}{32 \pi G_N \omega}\left[(d_i+3z-\theta-2)\frac{c_1-r^{-(d_i+3z-\theta-2)}}{c_1+r^{-(d_i+3z-\theta-2)}}-(d_i+2-z-\theta) \right]\\
    &\hspace*{7cm}+c_2\frac{r^{2(d_i+z-\theta)}}{(1+c_1 r^{d_i+3z-\theta-2})^2}+\mathcal{O}(\omega)\ .
\end{aligned}
\end{equation}
The response function $\chi_{a_i}$ at leading order exhibits non-trivial dependence on the radial coordinate and thus shows a very distinct behaviour compared to the response function $\chi_{xi}$.

\section{Discussion and Conclusion}\label{sec:4}

In this paper, we have studied and analysed the RG flow equations governing the shear response in hvLif theories from the holographic viewpoint. The presence of $U(1)$ gauge fields along with a dilaton complicate the analysis significantly since certain gauge field perturbations i.e. $a_i$ couples to the shear tensor modes $h_{xi}$ and $h_{ti}$. Focusing on the $q=0$ sector, our central observations are:
\begin{itemize}
	\item The shear viscosity at the leading order seems to saturate the KSS bound for all values of $z$ and $\theta$. Earlier works failed to make any statement about shear viscosity for $z>d_i+2-\theta$. This analysis gets around that issue of breakdown of hydrodynamic expansion for $z>d_i+2-\theta$. However, for the special value of $z=d_i-\theta$, we see a very interesting logarithmic correction at the first order. This does not violate the KSS bound but, necessitates the introduction of a UV cutoff to control potential divergences at the boundary. This particular logarithmic behaviour of the subleading correction to shear viscosity for $z=d_i-\theta$ seems to be a novel feature. Further, as discussed in previous works \cite{Balasubramanian:2010uk,Donos:2010tu}, dimensional reduction of null deformed $AdS_5$ results in $z=2$ Lifshitz theories (they have $\theta=0$) and is consistent with $z=d_i-\theta=2$. Given this observation, it is natural to ask if such logarithmic correction for $z=d_i-\theta$ can be explained from the perspective of the higher dimensional null deformed $AdS_5$ theory. It will be interesting to further explore the hydrodynamics of theories dual to such null deformed background.
	
	\item In the response function for $a_i$, we observe non-trivial flow even at leading order in $\chi_{a_i}$. We have performed the analysis in the near horizon and the near boundary region with appropriate approximations. In the near horizon region, the qualitative behaviour of $\chi_{a_i}$ seems to mimic that of $\chi_{xi}$. However, the near boundary analysis reveals a leading behaviour which scales as $\chi_{a_i} \sim r^{d_i+2-z-\theta}$. The response function happens to be convergent provided $z<d_i+2-\theta$. Thus, it seems this bound obtained in earlier works\cite{Kolekar:2016pnr,Kolekar:2016yzg,Mukherjee:2017ynv} can be interpreted as a regularity condition on the response function of the gauge field perturbations $a_i$. Earlier works constructed a linear combination involving all the perturbation modes which obfuscated the source of this constraint. Our present analysis seems to suggest that it is the gauge field perturbations exclusively which are responsible for the the constraint $z<d_i+2-\theta$.
\end{itemize}

\textbf{An aside on Markovianity index:} At this point one can ask for a more \emph{physical origin} for the constraints observed in this paper. In other words, we want to understand if the breakdown of the hydrodynamic expansion for a certain parameter range, namely $z>d_i+2-\theta$ has a more deeper origin or is simply a bug of these non-relativistic gravity duals. Towards that vein one perform a \emph{Markovianity index} analysis of the perturbations in the probe limit in the spirit of \cite{Ghosh:2020lel}. To be more elaborate, \cite{Ghosh:2020lel} studied probes couples to conserved currents in an AdS-Schwarzschild background. The effective coupling of the probe field is characterized by a single parameter, namely the \emph{Markovianity index} $\mathcal{M}$. Probe fields with $\mathcal{M}>-1$ exhibits short-lived memory and
behave analogous to the massive scalar probes. Probes with $\mathcal{M}\leq-1$, however, retain long-term memory. In the current context, the metric petrubations we study are coupled to conserved current i.e. the stress tensor.

More precisely, \cite{Ghosh:2020lel} starts from the effective action of a probe scalar of the form
\begin{equation}
    \cS_{eff}=-{1\over 2}\int d^{d+1}x \sqrt{-g}r^{d-1-\cM}\nabla^A \phi_\cM\nabla_A \phi_\cM+S_{bdy},
\end{equation}
describing a massless Klein-Gordon field minimally $\phi_\cM$ minimally coupled to gravity with metric being same as \eqref{hvmetric} and $\cM\in \mathbb{R}$ being some designer parameter modulating the coupling. To reiterate more concretely, With this designer scalar the central observation of \cite{Ghosh:2020lel} is that the scalar probe field $\phi_\cM$ is Markovian if $\cM>-1$ or else its non-Markovian. Written in terms of the usual Fourier modes the scalar field equation takes the form in the zero momentum limit\footnote{with $q\neq0$ the equation becomes $$\phi_\cM''+\left(-{\cM\over r}+{f'\over f}+i{2 \omega\over r f}\right)\phi_\cM'-\left({q^2\over f}+i{\omega \cM\over rf}\right)\phi_\cM=0.$$}
\begin{equation}\label{eq:probeom}
    \phi_\cM''+\left(-{\cM\over r}+{f'\over f}+i{2 \omega\over r f}\right)\phi_\cM'-i{\omega \cM\over r\,f}\,\phi_\cM=0
\end{equation}
Comparing \eqref{eq:probeom} with \eqref{nu-x-eqn} in the limit $\omega\to 0$ one can check that in this case the designer parameter turns out to be $$\cM=d_i+z-\theta-1.$$
Interestingly, this implies that constraining the perturbations to be Markovian also forces the probe to obey the null energy condition  \eqref{nullee}. In other words $$\cM>-1\quad\implies\quad d_i+z-\theta\geq 0.$$ The situation with \eqref{nu-t-eqn} or \eqref{chi-eqn} is much more complicated due to the coupling between the fields. One can simplify the situation by considering the near boundary region for \eqref{nu-t-eqn} where $f(r)\sim1$. In this regime, a comparison between \eqref{eq:probeom} and \eqref{nu-t-eqn} in the limit $\omega\to 0$ reveals\footnote{Near the boundary with $\omega\to 0$ and $q\to 0$ equation \eqref{nu-t-eqn} becomes  $$H_{ti}''+{z+\theta-d_i-1\over r}H_{ti}'=0,$$ where the term containing $a_i'$ gets dropped due to \eqref{eq:aiqneq0-soln}.} 
$$\cM=d_i-z-\theta+1.$$ Again imposing the Markovianity condition $\cM>-1$, we interestingly have $$z<d_i+2-\theta.$$ which is exactly the limit that we have obtained through the gauge field perturbations. Therefore the regularity condition of $z<d_i+2-\theta.$, analysed explicitly in the present analysis and observed earlier in \cite{Kolekar:2016pnr,Kolekar:2016yzg,Mukherjee:2017ynv} can also be attributed to the fact of the probes being Markovian. The above calculations although rudimentary seems to be hinting towards a connection between Markovianity index of the fluctuations and the breakdown of hydrodynamic expansion. An elaborate investigation of this issue is beyond the scope of this paper, which we hope to address in future works.

\paragraph{}Our strategy of analysing the near horizon and near boundary regions separately opens up some possible new directions in the hydrodynamics of hvLif theories. Our analysis is restrictive in the sense that we analysed the $q=0$ sector only. A natural extension will be to understand the $q \neq 0$ sector and check if one recovers any new transport coefficient at linear order in $q$. Another interesting question will be to explore if regularity conditions imposed on response functions of higher order transport coefficients leads to any further constraint on the Lifshitz exponent $z$. We are looking forward to analysing the flow equations both analytically and numerically to comment on higher order transport coefficients which are yet unexplored in the literature. We subsequently plan on studying the RG flow of response functions that arise in the sound channel and scalar channel.

\vspace{5mm}

\section*{Acknowledgements:} We would like to thank K. Narayan for his detailed feedback which significantly improved this manuscript. We are also thankful to Nabamita Banerjee, Suvankar Dutta and A. Sivakumar for valuable inputs. DM acknowledges the support and hospitality of IISER Bhopal during the course of this work. The work of AC is supported in part by the South African Research Chairs initiative of the National Research Foundation, grant number 78554. The work of DM is supported in part by the grants CRG/2018/002373 and SB/SJF/2019-20/08.

\appendix
\section{Reviewing hyperscaling violating Lifshitz spacetimes}
\label{hvlif-review}

The metric \eqref{hvmetric} is a solution to the Einstein-Maxwell-dilaton
action
\begin{equation}
\label{hvaction}
S= -\frac{1}{16 \pi G_{N}^{(d+1)}}\int d^{d+1}x\  \sqrt{-G}\left[R-\frac{1}{2}\partial_{\mu}\phi\partial^{\mu}\phi-\frac{Z(\phi)}{4}F_{\mu \nu}F^{\mu \nu}+V(\phi) \right]\ ,
\end{equation}
where the various fields and parameters appearing in the action are listed as follows:
\begin{eqnarray}
\phi&=&\sqrt{2(d_i-\theta)(z-\theta/d_i-1)}\ \log r\ ,\\
\label{bg-gauge-fld}
A_t&=&\frac{\alpha f(r)}{r^{d_i+z-\theta}}\ , \qquad
\alpha=-\sqrt{\frac{2(z-1)}{d_i+z-\theta}}\ , \qquad A_i=0\ .\\
V(\phi)&=&(d_i+z-\theta)(d_i+z-\theta-1)r^{-\frac{2\theta}{d_i}}\ ; \quad 
Z(\phi)=r^{\frac{2\theta}{d_i}+2d_i-2\theta}=e^{\lambda \phi} . \label{hvaction-parameters}
\end{eqnarray}
The null energy conditions following from (\ref{hvmetric}) give constraints
on the Lifshitz $z$ and hyperscaling violating $\theta$ exponents\ 
\be\label{nullee}
(z-1)(d_i+z-\theta)\geq 0\ ,\qquad (d_i-\theta)(d_i(z-1)-\theta)\geq 0\ .
\ee
Varying with $\gbar_{\mu \nu}$, $\Abar_{\mu}$ and $\bar{\phi}$, we obtain the following equations of motion,
\begin{equation}\label{einsteineqn}
\bar{R}_{\mu\nu}=\frac{1}{2}\partial_{\mu}\bar{\phi}\partial_{\nu}\bar{\phi} -\gbar_{\mu\nu}\frac{V(\bar{\phi})}{d-1} + \frac{Z(\bar{\phi})}{2}\gbar^{\rho\sigma}\Fbar_{\rho\mu}\Fbar_{\sigma\nu}
- \frac{Z(\bar{\phi})}{4(d-1)}\gbar_{\mu\nu}\Fbar_{\rho\sigma}\Fbar^{\rho\sigma}\ ,
\end{equation}
\begin{equation}\label{gaugeeqn}
\nabla_{\mu}(Z(\bar{\phi})\Fbar^{\mu\nu})=0\ ,
\end{equation}
\begin{equation}\label{scalareqn}
\frac{1}{\sqrt{-\gbar}}\partial_{\mu}(\sqrt{-\gbar}\gbar^{\mu\nu}\partial_{\nu}\bar{\phi})+\frac{\partial V(\bar{\phi})}{\partial\bar{\phi}}-\frac{1}{4}\frac{\partial Z(\bar{\phi})}{\partial\bar{\phi}}\Fbar_{\rho\sigma}\Fbar^{\rho\sigma}=0\ .
\end{equation}
Note that from \eqref{einsteineqn} it follows that:
\begin{equation}
\bar{R}=\bar{R}_{\mu \nu}\gbar^{\mu \nu}=\frac{1}{2}\partial_{\rho}\bar{\phi} \partial^{\rho}\bar{\phi}-\frac{d+1}{d-1}V(\bar{\phi})+\frac{Z(\bar{\phi})}{2}\gbar^{\rho \sigma}\Fbar_{\rho}^{\ \lambda}\Fbar_{\sigma \lambda}-\frac{Z(\bar{\phi})(d+1)}{4(d-1)}\Fbar^2
\end{equation}
Alternatively, we can write \eqref{einsteineqn} as:
\begin{equation}
\label{recast-einstein-eqn}
\begin{aligned}
\bar{R}_{\mu \nu}-\frac{1}{2}\gbar_{\mu \nu}\bar{R}&=\frac{1}{2}\partial_{\mu}\bar{\phi}\partial_{\nu}\bar{\phi} -\gbar_{\mu\nu}\frac{V(\bar{\phi})}{d-1} + \frac{Z(\bar{\phi})}{2}\gbar^{\rho\sigma}\Fbar_{\rho\mu}\Fbar_{\sigma\nu}
- \frac{Z(\bar{\phi})}{4(d-1)}\gbar_{\mu\nu}\Fbar_{\rho\sigma}\Fbar^{\rho\sigma}\\
&\hspace*{0.5cm}-\frac{1}{2}\gbar_{\mu \nu}\left[\frac{1}{2}\partial_{\rho}\bar{\phi} \partial^{\rho}\bar{\phi}-\frac{d+1}{d-1}V(\bar{\phi})+\frac{Z(\bar{\phi})}{2}\gbar^{\rho \sigma}\Fbar_{\rho}^{\ \lambda}\Fbar_{\sigma \lambda}-\frac{Z(\bar{\phi})(d+1)}{4(d-1)}\Fbar^2\right]\\
\Rightarrow\ \bar{R}_{\mu \nu}-\frac{1}{2}\gbar_{\mu \nu}\bar{R}&=\frac{1}{2}\partial_{\mu}\bar{\phi}\partial_{\nu}\bar{\phi}-\frac{1}{4}\gbar_{\mu \nu}\partial_{\rho}\bar{\phi}\partial^{\rho}\bar{\phi}+\frac{Z(\bar{\phi})}{2}\Fbar_{\rho \mu}\Fbar^{\rho}_{\ \nu}-\frac{Z(\bar{\phi})}{8}\gbar_{\mu \nu}\Fbar^2+\frac{V(\bar{\phi})}{2}\gbar_{\mu \nu}
\end{aligned}
\end{equation}

\section{Perturbations to hvLif spacetimes}
\label{app-b}

The perturbed action up to second order terms is given by
\begin{equation}
\label{gen-action}
\begin{aligned}
S^{(2)}&=\frac{-1}{16 \pi G_N}\int dr\ d^{d}k \left[\mathcal{A}(r)h''_{ti}h_{ti}+\tilde{\mathcal{A}}(r)h''_{xi}h_{xi}+\mathcal{B}(r)h^{'2}_{ti}+\tilde{\mathcal{B}}(r)h^{'2}_{xi} +\mathcal{C}(r)h_{ti}h'_{ti}\right.\\
&\hspace{0.1cm}+\tilde{\mathcal{C}}(r)h_{xi}h'_{xi}+\mathcal{D}(r)h_{ti}^2+\tilde{\mathcal{D}}(r)h_{xi}^2-g(r)h_{ti}a'_i+\mathcal{H}(r,q)h_{ti}^2+\tilde{\mathcal{H}}(r,\omega)h_{xi}^2\\
&\hspace{5cm}\left.+2\mathcal{J}(r,q,\omega)h_{ti}h_{xi}+\mathcal{M}(r){a'}^{2}_i+\mathcal{N}(r,q,\omega)a_i^2\right]\ ,
\end{aligned}
\end{equation}
where the various functions appearing in the action is given by:
\begin{equation*}
\begin{aligned}
&\mathcal{A}(r)=-2r^{3-d_i+z+\frac{\theta}{d_i}(d_i-4)} \quad;\  \tilde{\mathcal{A}}(r)=2fr^{5-d_i-z+\frac{\theta}{d_i}(d_i-4)}&\\
&\mathcal{B}(r)=-\frac{3}{2}r^{3-d_i+z+\frac{\theta}{d_i}(d_i-4)}\quad ;\ \tilde{\mathcal{B}}(r)=\frac{3}{2}fr^{5-d_i-z+\frac{\theta}{d_i}(d_i-4)}&\\
&\mathcal{C}(r)=\left[3d_i-8-3z+\frac{12\theta}{d_i}-3\theta-\frac{d_i+z-\theta}{f}\right]r^{2-d_i+z+\frac{\theta}{d_i}(d_i-4)}&\\
&\tilde{\mathcal{C}}(r)=\left[-2d_i-2z+2\theta+14f-\frac{12\theta}{d_i}f\right]r^{4-d_i-z+\frac{\theta}{d_i}(d_i-4)}&\\
&\mathcal{D}(r)=\left[-2+3d_i-\frac{d_i^2}{2}+(d_i-3)z-\frac{z^2}{2}+\frac{5-d_i}{d_i}z\theta+\left(-8+\frac{10}{d_i}+d_i\right)\theta\right.&\\
&\hspace{1.4cm}+\left(\frac{5}{d_i}-\frac{10}{d_i^2}-\frac{1}{2}\right)\theta^2+\frac{1}{f}\left((d_i-1)(d_i+z)+\frac{2-d_i}{d_i}z\theta-\frac{2-d_i}{d_i}\theta^2+(3-2d_i)\theta\right)&\\
&\hspace{1.4cm}\left. +\frac{1}{f^2}\left(-\frac{d_i^2}{2}-zd_i-\frac{z^2}{2}+z\theta+d_i\theta-\frac{\theta^2}{2}\right)\right]r^{1-d_i+z+\frac{\theta}{d_i}(d_i-4)}&\\
&\tilde{\mathcal{D}}(r)=\left[-3d_i-3z+6\theta+\frac{3z\theta}{d_i}-\frac{3\theta^2}{d_i}+10f-\frac{20\theta}{d_i}f+\frac{10\theta^2}{d_i^2}f\right]r^{3-d_i-z+\frac{\theta}{d_i}(d_i-4)}&
\end{aligned}
\end{equation*}
\begin{equation*}
\begin{aligned}
&g(r)=(d_i+z-\theta)\alpha r^{2-\frac{2\theta}{d_i}}\ ;\  \mathcal{H}(r,q,\omega)=\frac{q^2}{2}\frac{r^{3-d_i+z+\frac{\theta}{d_i}(d_i-4)}}{f}\ ;\ \tilde{\mathcal{H}}(r,q,\omega)=\frac{\omega^2}{2}\frac{r^{3-d_i+z+\frac{\theta}{d_i}(d_i-4)}}{f}&\\
&\mathcal{J}(r,q,\omega)=\frac{\omega q}{2}\frac{r^{3-d_i+z+\frac{\theta}{d_i}(d_i-4)}}{f}\quad ;\ \mathcal{M}(r)=-\frac{1}{2}fr^{d_i+3-z-\theta}&\\
&\mathcal{N}(r,q,\omega)=\frac{\omega^2}{2}\frac{r^{d_i+1+z-\theta}}{f}-\frac{q^2}{2}r^{d_i+3-z-\theta}&
\end{aligned}
\end{equation*}
The modes $h_{ti}(t,r,x), h_{xi}(t,r,x)$ and $a_i(t,r,x)$ form a decoupled set of equations along with a constraint equation which can be solved perturbatively for every $x_i \in \{x_2,...,x_{d_i}\}$ and $x \equiv x_1$.

\section{Solution of $\zeta_{a_i}$ at second order near boundary}\label{app:zetaabdy-soln}

As demonstrated earlier, the equation governing the second order correction to $\zeta_{a_i}$ near the boundary is given by \eqref{eq:zetabdy2eqn}. The generic solution to this equation is given by
\begin{equation}
\begin{aligned}
    \bzeta^{(2)}(r)&=c_3 \frac{r^{2(d_i+z-\theta)}}{(1+c_1 r^{d_i+3z-\theta-2})^2}-i\frac{5r^{d_i+z-\theta+2}}{16 \pi G_N(d_i+z-\theta-2)(1+c_1 r^{d_i+3z-\theta-2})^2}\\
    &-ic_1\frac{r^{2(d_i+2z-\theta)}((d_i+5z-\theta-2)+c_1z r^{d_i+3z-\theta-2})}{16 \pi G_N\ z(d_i+5z-\theta-2)(1+c_1 r^{d_i+3z-\theta-2})^2}\\
    &+i\frac{r^{2(d_i+z-\theta)}\left[3r^2-3c_1^2 r^{2(d_i+3z-\theta-1)}+128\pi^2 G_N^2\ c_2^2 r^{2(d_i+2z-\theta-1)} \right]}{8\pi G_N (d_i+3z-\theta-2)(1+c_1 r^{d_i+3z-\theta-2})^3}\\
    &+\frac{6i}{8 \pi G_N}\frac{zr^{d_i+z-\theta+2}\ _2F_1[1,\frac{2+\theta-d_i-z}{d_i+3z-\theta-2};\frac{2z}{d_i+3z-\theta-2};-c_1r^{d_i+3z-\theta-2}]}{(d_i+z-\theta-2)(d_i+3z-\theta-2)(1+c_2 r^{d_i+3z-\theta-2})^2}\\
    &+\frac{3i}{8\pi G_N}\frac{2zc_1^2}{(d_i+3z-\theta-2)(d_i+5z-\theta-2)}\frac{r^{3d_i+7z-3\theta-2}}{(1+c_1r^{d_i+3z-\theta-2})^2}\times\\
    &\hspace*{3cm}_2F_1\left[ 1,\frac{d_i+5z-\theta-2}{d_i+3z-\theta-2};\frac{2(d_i+4z-\theta-2)}{d_i+3z-\theta-2};-c_1r^{d_i+3z-\theta-2}\right]\ ,
\end{aligned}
\end{equation}
which is valid when $d_i+z-\theta \neq 2$. 

When $d_i+z-\theta=2$ and $z\neq 2$, we have the solution
\begin{equation}
     \bzeta^{(2)}(r)=\frac{r^4 \left(-\frac{256 i \pi G_N c_2^2 }{c_1^2 z r^{2 z}+c_1 z}-\frac{i r^{2 z-4} \left(c_1 r^{2 z} \left(\frac{c_1 r^{2 z}}{3 z-2}+\frac{1}{z-1}\right)+\frac{1}{z-2}\right)}{\pi  G_N}+32 \kappa_1\right)}{32 \left(\text{c1} r^{2 z}+1\right)^2}\ .
\end{equation}
When $z=2$ and $d_i=\theta$, we recover the solution
\begin{equation}
     \bzeta^{(2)}(r)=\frac{r^4 \left(16 \kappa_2-\frac{i \left(\frac{64 \pi ^2 G_N^2 c_2^2}{c_1^2 r^4+c_1}+\frac{c_1^2 r^8}{8}+\frac{c_1 r^4}{2}+\log (r)\right)}{\pi  G_N}\right)}{16 \left(c_1 r^4+1\right)^2}.
\end{equation}
All of them vanish in the limit $r \rightarrow 0$ (near boundary) thus leaving the constants $c_3, \kappa_1$ and $\kappa_2$ unfixed.

\bibliographystyle{JHEP}
\bibliography{Hydro.bib}

\providecommand{\href}[2]{#2}\begingroup\raggedright\begin{thebibliography}{10}

\bibitem{Maldacena:1997re}
J.~M. Maldacena, \emph{{The Large N limit of superconformal field theories and
  supergravity}}, \href{https://doi.org/10.1023/A:1026654312961}{\emph{Adv.
  Theor. Math. Phys.} {\bfseries 2} (1998) 231}
  [\href{https://arxiv.org/abs/hep-th/9711200}{{\ttfamily hep-th/9711200}}].

\bibitem{Gubser:1998bc}
S.~S. Gubser, I.~R. Klebanov and A.~M. Polyakov, \emph{{Gauge theory
  correlators from noncritical string theory}},
  \href{https://doi.org/10.1016/S0370-2693(98)00377-3}{\emph{Phys. Lett. B}
  {\bfseries 428} (1998) 105}
  [\href{https://arxiv.org/abs/hep-th/9802109}{{\ttfamily hep-th/9802109}}].

\bibitem{Witten:1998qj}
E.~Witten, \emph{{Anti-de Sitter space and holography}},
  \href{https://doi.org/10.4310/ATMP.1998.v2.n2.a2}{\emph{Adv. Theor. Math.
  Phys.} {\bfseries 2} (1998) 253}
  [\href{https://arxiv.org/abs/hep-th/9802150}{{\ttfamily hep-th/9802150}}].

\bibitem{Aharony:1999ti}
O.~Aharony, S.~S. Gubser, J.~M. Maldacena, H.~Ooguri and Y.~Oz, \emph{{Large N
  field theories, string theory and gravity}},
  \href{https://doi.org/10.1016/S0370-1573(99)00083-6}{\emph{Phys. Rept.}
  {\bfseries 323} (2000) 183}
  [\href{https://arxiv.org/abs/hep-th/9905111}{{\ttfamily hep-th/9905111}}].

\bibitem{Gubser:2009qt}
S.~S. Gubser and F.~D. Rocha, \emph{{Peculiar properties of a charged dilatonic
  black hole in $AdS_{5}$}},
  \href{https://doi.org/10.1103/PhysRevD.81.046001}{\emph{Phys. Rev. D}
  {\bfseries 81} (2010) 046001}
  [\href{https://arxiv.org/abs/0911.2898}{{\ttfamily 0911.2898}}].

\bibitem{Cadoni:2009xm}
M.~Cadoni, G.~D'Appollonio and P.~Pani, \emph{{Phase transitions between
  Reissner-Nordstrom and dilatonic black holes in 4D AdS spacetime}},
  \href{https://doi.org/10.1007/JHEP03(2010)100}{\emph{JHEP} {\bfseries 03}
  (2010) 100} [\href{https://arxiv.org/abs/0912.3520}{{\ttfamily 0912.3520}}].

\bibitem{Goldstein:2009cv}
K.~Goldstein, S.~Kachru, S.~Prakash and S.~P. Trivedi, \emph{{Holography of
  Charged Dilaton Black Holes}},
  \href{https://doi.org/10.1007/JHEP08(2010)078}{\emph{JHEP} {\bfseries 08}
  (2010) 078} [\href{https://arxiv.org/abs/0911.3586}{{\ttfamily 0911.3586}}].

\bibitem{Charmousis:2010zz}
C.~Charmousis, B.~Gouteraux, B.~S. Kim, E.~Kiritsis and R.~Meyer,
  \emph{{Effective Holographic Theories for low-temperature condensed matter
  systems}}, \href{https://doi.org/10.1007/JHEP11(2010)151}{\emph{JHEP}
  {\bfseries 11} (2010) 151} [\href{https://arxiv.org/abs/1005.4690}{{\ttfamily
  1005.4690}}].

\bibitem{Perlmutter:2010qu}
E.~Perlmutter, \emph{{Domain Wall Holography for Finite Temperature Scaling
  Solutions}}, \href{https://doi.org/10.1007/JHEP02(2011)013}{\emph{JHEP}
  {\bfseries 02} (2011) 013} [\href{https://arxiv.org/abs/1006.2124}{{\ttfamily
  1006.2124}}].

\bibitem{Bertoldi:2010ca}
G.~Bertoldi, B.~A. Burrington and A.~W. Peet, \emph{{Thermal behavior of
  charged dilatonic black branes in AdS and UV completions of Lifshitz-like
  geometries}}, \href{https://doi.org/10.1103/PhysRevD.82.106013}{\emph{Phys.
  Rev. D} {\bfseries 82} (2010) 106013}
  [\href{https://arxiv.org/abs/1007.1464}{{\ttfamily 1007.1464}}].

\bibitem{Kim:2010zq}
B.~S. Kim, E.~Kiritsis and C.~Panagopoulos, \emph{{Holographic quantum
  criticality and strange metal transport}},
  \href{https://doi.org/10.1088/1367-2630/14/4/043045}{\emph{New J. Phys.}
  {\bfseries 14} (2012) 043045}
  [\href{https://arxiv.org/abs/1012.3464}{{\ttfamily 1012.3464}}].

\bibitem{Iizuka:2011hg}
N.~Iizuka, N.~Kundu, P.~Narayan and S.~P. Trivedi, \emph{{Holographic Fermi and
  Non-Fermi Liquids with Transitions in Dilaton Gravity}},
  \href{https://doi.org/10.1007/JHEP01(2012)094}{\emph{JHEP} {\bfseries 01}
  (2012) 094} [\href{https://arxiv.org/abs/1105.1162}{{\ttfamily 1105.1162}}].

\bibitem{Ogawa:2011bz}
N.~Ogawa, T.~Takayanagi and T.~Ugajin, \emph{{Holographic Fermi Surfaces and
  Entanglement Entropy}},
  \href{https://doi.org/10.1007/JHEP01(2012)125}{\emph{JHEP} {\bfseries 01}
  (2012) 125} [\href{https://arxiv.org/abs/1111.1023}{{\ttfamily 1111.1023}}].

\bibitem{Cremonini:2011ej}
S.~Cremonini and P.~Szepietowski, \emph{{Generating Temperature Flow for eta/s
  with Higher Derivatives: From Lifshitz to AdS}},
  \href{https://doi.org/10.1007/JHEP02(2012)038}{\emph{JHEP} {\bfseries 02}
  (2012) 038} [\href{https://arxiv.org/abs/1111.5623}{{\ttfamily 1111.5623}}].

\bibitem{Huijse:2011ef}
L.~Huijse, S.~Sachdev and B.~Swingle, \emph{{Hidden Fermi surfaces in
  compressible states of gauge-gravity duality}},
  \href{https://doi.org/10.1103/PhysRevB.85.035121}{\emph{Phys. Rev. B}
  {\bfseries 85} (2012) 035121}
  [\href{https://arxiv.org/abs/1112.0573}{{\ttfamily 1112.0573}}].

\bibitem{Dong:2012se}
X.~Dong, S.~Harrison, S.~Kachru, G.~Torroba and H.~Wang, \emph{{Aspects of
  holography for theories with hyperscaling violation}},
  \href{https://doi.org/10.1007/JHEP06(2012)041}{\emph{JHEP} {\bfseries 06}
  (2012) 041} [\href{https://arxiv.org/abs/1201.1905}{{\ttfamily 1201.1905}}].

\bibitem{Kiritsis:2012ta}
E.~Kiritsis, \emph{{Lorentz violation, Gravity, Dissipation and Holography}},
  \href{https://doi.org/10.1007/JHEP01(2013)030}{\emph{JHEP} {\bfseries 01}
  (2013) 030} [\href{https://arxiv.org/abs/1207.2325}{{\ttfamily 1207.2325}}].

\bibitem{Bhattacharya:2012zu}
J.~Bhattacharya, S.~Cremonini and A.~Sinkovics, \emph{{On the IR completion of
  geometries with hyperscaling violation}},
  \href{https://doi.org/10.1007/JHEP02(2013)147}{\emph{JHEP} {\bfseries 02}
  (2013) 147} [\href{https://arxiv.org/abs/1208.1752}{{\ttfamily 1208.1752}}].

\bibitem{Alishahiha:2012cm}
M.~Alishahiha and H.~Yavartanoo, \emph{{On Holography with Hyperscaling
  Violation}}, \href{https://doi.org/10.1007/JHEP11(2012)034}{\emph{JHEP}
  {\bfseries 11} (2012) 034} [\href{https://arxiv.org/abs/1208.6197}{{\ttfamily
  1208.6197}}].

\bibitem{Hoyos:2013qna}
C.~Hoyos, B.~S. Kim and Y.~Oz, \emph{{Lifshitz Field Theories at Non-Zero
  Temperature, Hydrodynamics and Gravity}},
  \href{https://doi.org/10.1007/JHEP03(2014)029}{\emph{JHEP} {\bfseries 03}
  (2014) 029} [\href{https://arxiv.org/abs/1309.6794}{{\ttfamily 1309.6794}}].

\bibitem{Sadeghi:2014zia}
J.~Sadeghi and A.~Asadi, \emph{{Hydrodynamics in a black brane with
  hyperscaling violation metric background}},
  \href{https://doi.org/10.1139/cjp-2014-0067}{\emph{Can. J. Phys.} {\bfseries
  92} (2014) 1570} [\href{https://arxiv.org/abs/1404.5282}{{\ttfamily
  1404.5282}}].

\bibitem{Ghodrati:2014spa}
M.~Ghodrati, \emph{{Hyperscaling Violating Solution in Coupled Dilaton-Squared
  Curvature Gravity}},
  \href{https://doi.org/10.1103/PhysRevD.90.044055}{\emph{Phys. Rev. D}
  {\bfseries 90} (2014) 044055}
  [\href{https://arxiv.org/abs/1404.5399}{{\ttfamily 1404.5399}}].

\bibitem{Kiritsis:2015doa}
E.~Kiritsis and Y.~Matsuo, \emph{{Charge-hyperscaling violating Lifshitz
  hydrodynamics from black-holes}},
  \href{https://doi.org/10.1007/JHEP12(2015)076}{\emph{JHEP} {\bfseries 12}
  (2015) 076} [\href{https://arxiv.org/abs/1508.02494}{{\ttfamily
  1508.02494}}].

\bibitem{Kuang:2015mlf}
X.-M. Kuang, J.-P. Wu and J.-P. Wu, \emph{{Analytical shear viscosity in
  hyperscaling violating black brane}},
  \href{https://doi.org/10.1016/j.physletb.2017.08.060}{\emph{Phys. Lett. B}
  {\bfseries 773} (2017) 422}
  [\href{https://arxiv.org/abs/1511.03008}{{\ttfamily 1511.03008}}].

\bibitem{Taylor:2015glc}
M.~Taylor, \emph{{Lifshitz holography}},
  \href{https://doi.org/10.1088/0264-9381/33/3/033001}{\emph{Class. Quant.
  Grav.} {\bfseries 33} (2016 [a ]) 033001}
  [\href{https://arxiv.org/abs/1512.03554}{{\ttfamily 1512.03554}}].

\bibitem{Kolekar:2016pnr}
K.~S. Kolekar, D.~Mukherjee and K.~Narayan, \emph{{Hyperscaling violation and
  the shear diffusion constant}},
  \href{https://doi.org/10.1016/j.physletb.2016.06.046}{\emph{Phys. Lett. B}
  {\bfseries 760} (2016) 86}
  [\href{https://arxiv.org/abs/1604.05092}{{\ttfamily 1604.05092}}].

\bibitem{Kiritsis:2016rcb}
E.~Kiritsis and Y.~Matsuo, \emph{{Hyperscaling-Violating Lifshitz hydrodynamics
  from black-holes: Part II}},
  \href{https://doi.org/10.1007/JHEP03(2017)041}{\emph{JHEP} {\bfseries 03}
  (2017) 041} [\href{https://arxiv.org/abs/1611.04773}{{\ttfamily
  1611.04773}}].

\bibitem{Kolekar:2016yzg}
K.~S. Kolekar, D.~Mukherjee and K.~Narayan, \emph{{Notes on hyperscaling
  violating Lifshitz and shear diffusion}},
  \href{https://doi.org/10.1103/PhysRevD.96.026003}{\emph{Phys. Rev. D}
  {\bfseries 96} (2017) 026003}
  [\href{https://arxiv.org/abs/1612.05950}{{\ttfamily 1612.05950}}].

\bibitem{Ling:2016ien}
Y.~Ling, Z.-Y. Xian and Z.~Zhou, \emph{{Holographic Shear Viscosity in
  Hyperscaling Violating Theories without Translational Invariance}},
  \href{https://doi.org/10.1007/JHEP11(2016)007}{\emph{JHEP} {\bfseries 11}
  (2016) 007} [\href{https://arxiv.org/abs/1605.03879}{{\ttfamily
  1605.03879}}].

\bibitem{Ling:2016yxy}
Y.~Ling, Z.~Xian and Z.~Zhou, \emph{{Power Law of Shear Viscosity in
  Einstein-Maxwell-Dilaton-Axion model}},
  \href{https://doi.org/10.1088/1674-1137/41/2/023104}{\emph{Chin. Phys. C}
  {\bfseries 41} (2017) 023104}
  [\href{https://arxiv.org/abs/1610.08823}{{\ttfamily 1610.08823}}].

\bibitem{Davison:2016auk}
R.~A. Davison, S.~Grozdanov, S.~Janiszewski and M.~Kaminski, \emph{{Momentum
  and charge transport in non-relativistic holographic fluids from Ho\v{r}ava
  gravity}}, \href{https://doi.org/10.1007/JHEP11(2016)170}{\emph{JHEP}
  {\bfseries 11} (2016) 170}
  [\href{https://arxiv.org/abs/1606.06747}{{\ttfamily 1606.06747}}].

\bibitem{Hartong:2016nyx}
J.~Hartong, N.~A. Obers and M.~Sanchioni, \emph{{Lifshitz Hydrodynamics from
  Lifshitz Black Branes with Linear Momentum}},
  \href{https://doi.org/10.1007/JHEP10(2016)120}{\emph{JHEP} {\bfseries 10}
  (2016) 120} [\href{https://arxiv.org/abs/1606.09543}{{\ttfamily
  1606.09543}}].

\bibitem{Eberlein:2016luy}
A.~Eberlein, A.~A. Patel and S.~Sachdev, \emph{{Shear viscosity at the
  Ising-nematic quantum critical point in two dimensional metals}},
  \href{https://doi.org/10.1103/PhysRevB.95.075127}{\emph{Phys. Rev. B}
  {\bfseries 95} (2017) 075127}
  [\href{https://arxiv.org/abs/1607.03894}{{\ttfamily 1607.03894}}].

\bibitem{Mukherjee:2017ynv}
D.~Mukherjee and K.~Narayan, \emph{{Hyperscaling violation, quasinormal modes
  and shear diffusion}},
  \href{https://doi.org/10.1007/JHEP12(2017)023}{\emph{JHEP} {\bfseries 12}
  (2017) 023} [\href{https://arxiv.org/abs/1707.07490}{{\ttfamily
  1707.07490}}].

\bibitem{Hartnoll:2018xxg}
S.~A. Hartnoll, A.~Lucas and S.~Sachdev, \emph{{Holographic Quantum Matter}}.
  MIT Press, 2018 [a shorter version is available at
  \href{https://arxiv.org/pdf/1612.07324.pdf}{arXiv:1612.07324}].

\bibitem{Herrera-Aguilar:2021top}
A.~Herrera-Aguilar, J.~A. Herrera-Mendoza and D.~F. Higuita-Borja,
  \emph{{Rotating spacetimes generalizing Lifshitz black holes}},
  \href{https://doi.org/10.1140/epjc/s10052-021-09682-9}{\emph{Eur. Phys. J. C}
  {\bfseries 81} (2021) 874}
  [\href{https://arxiv.org/abs/2104.14514}{{\ttfamily 2104.14514}}].

\bibitem{Yuan:2020fvv}
H.~Yuan and X.-H. Ge, \emph{{Pole-skipping and hydrodynamic analysis in
  Lifshitz, AdS$_{2}$ and Rindler geometries}},
  \href{https://doi.org/10.1007/JHEP06(2021)165}{\emph{JHEP} {\bfseries 06}
  (2021) 165} [\href{https://arxiv.org/abs/2012.15396}{{\ttfamily
  2012.15396}}].

\bibitem{Narayan:2012hk}
K.~Narayan, \emph{{On Lifshitz scaling and hyperscaling violation in string
  theory}}, \href{https://doi.org/10.1103/PhysRevD.85.106006}{\emph{Phys. Rev.
  D} {\bfseries 85} (2012) 106006}
  [\href{https://arxiv.org/abs/1202.5935}{{\ttfamily 1202.5935}}].

\bibitem{Singh:2012un}
H.~Singh, \emph{{Lifshitz/Schr\"odinger Dp-branes and dynamical exponents}},
  \href{https://doi.org/10.1007/JHEP07(2012)082}{\emph{JHEP} {\bfseries 07}
  (2012) 082} [\href{https://arxiv.org/abs/1202.6533}{{\ttfamily 1202.6533}}].

\bibitem{Balasubramanian:2010uk}
K.~Balasubramanian and K.~Narayan, \emph{{Lifshitz spacetimes from AdS null and
  cosmological solutions}},
  \href{https://doi.org/10.1007/JHEP08(2010)014}{\emph{JHEP} {\bfseries 08}
  (2010) 014} [\href{https://arxiv.org/abs/1005.3291}{{\ttfamily 1005.3291}}].

\bibitem{Donos:2010tu}
A.~Donos and J.~P. Gauntlett, \emph{{Lifshitz Solutions of D=10 and D=11
  supergravity}}, \href{https://doi.org/10.1007/JHEP12(2010)002}{\emph{JHEP}
  {\bfseries 12} (2010) 002} [\href{https://arxiv.org/abs/1008.2062}{{\ttfamily
  1008.2062}}].

\bibitem{Ross:2011gu}
S.~F. Ross, \emph{{Holography for asymptotically locally Lifshitz spacetimes}},
  \href{https://doi.org/10.1088/0264-9381/28/21/215019}{\emph{Class. Quant.
  Grav.} {\bfseries 28} (2011) 215019}
  [\href{https://arxiv.org/abs/1107.4451}{{\ttfamily 1107.4451}}].

\bibitem{Christensen:2013rfa}
M.~H. Christensen, J.~Hartong, N.~A. Obers and B.~Rollier, \emph{{Boundary
  Stress-Energy Tensor and Newton-Cartan Geometry in Lifshitz Holography}},
  \href{https://doi.org/10.1007/JHEP01(2014)057}{\emph{JHEP} {\bfseries 01}
  (2014) 057} [\href{https://arxiv.org/abs/1311.6471}{{\ttfamily 1311.6471}}].

\bibitem{Chemissany:2014xsa}
W.~Chemissany and I.~Papadimitriou, \emph{{Lifshitz holography: The whole
  shebang}}, \href{https://doi.org/10.1007/JHEP01(2015)052}{\emph{JHEP}
  {\bfseries 01} (2015) 052} [\href{https://arxiv.org/abs/1408.0795}{{\ttfamily
  1408.0795}}].

\bibitem{Hartong:2015wxa}
J.~Hartong, E.~Kiritsis and N.~A. Obers, \emph{{Field Theory on Newton-Cartan
  Backgrounds and Symmetries of the Lifshitz Vacuum}},
  \href{https://doi.org/10.1007/JHEP08(2015)006}{\emph{JHEP} {\bfseries 08}
  (2015) 006} [\href{https://arxiv.org/abs/1502.00228}{{\ttfamily
  1502.00228}}].

\bibitem{Kovtun:2004de}
P.~Kovtun, D.~T. Son and A.~O. Starinets, \emph{{Viscosity in strongly
  interacting quantum field theories from black hole physics}},
  \href{https://doi.org/10.1103/PhysRevLett.94.111601}{\emph{Phys. Rev. Lett.}
  {\bfseries 94} (2005) 111601}
  [\href{https://arxiv.org/abs/hep-th/0405231}{{\ttfamily hep-th/0405231}}].

\bibitem{Iqbal:2008by}
N.~Iqbal and H.~Liu, \emph{{Universality of the hydrodynamic limit in AdS/CFT
  and the membrane paradigm}},
  \href{https://doi.org/10.1103/PhysRevD.79.025023}{\emph{Phys. Rev. D}
  {\bfseries 79} (2009) 025023}
  [\href{https://arxiv.org/abs/0809.3808}{{\ttfamily 0809.3808}}].

\bibitem{Mamo:2012sy}
K.~A. Mamo, \emph{{Holographic RG flow of the shear viscosity to entropy
  density ratio in strongly coupled anisotropic plasma}},
  \href{https://doi.org/10.1007/JHEP10(2012)070}{\emph{JHEP} {\bfseries 10}
  (2012) 070} [\href{https://arxiv.org/abs/1205.1797}{{\ttfamily 1205.1797}}].

\bibitem{Ghosh:2020lel}
J.~K. Ghosh, R.~Loganayagam, S.~G. Prabhu, M.~Rangamani, A.~Sivakumar and
  V.~Vishal, \emph{{Effective field theory of stochastic diffusion from
  gravity}}, \href{https://doi.org/10.1007/JHEP05(2021)130}{\emph{JHEP}
  {\bfseries 05} (2021) 130}
  [\href{https://arxiv.org/abs/2012.03999}{{\ttfamily 2012.03999}}].

\end{thebibliography}\endgroup

\end{document}